\documentclass[pra,aps,10pt,superscriptaddress,twocolumn,floatfix,showpacs]{revtex4-1}
\usepackage{graphicx}
\usepackage{amsmath}
\usepackage{amsfonts}
\usepackage{amssymb}
\usepackage{epsfig}
\usepackage[pdftex]{color}
\usepackage{amsmath,graphicx,amssymb,xcolor,subfigure,upgreek}
\usepackage[colorlinks, linkcolor=blue, citecolor=blue, urlcolor=blue, breaklinks=true]{hyperref}
\usepackage{bbm}
\usepackage{color}

\newcommand{\avg}[1]{\left\langle#1\right\rangle}

\newcommand{\braket}[2]{\langle{#1}|{#2}\rangle}

\renewcommand{\max}[0]{\mathrm{max}}

\renewcommand{\emph}[1]{{\it #1}}
\renewcommand{\vec}[1]{\boldsymbol{#1}}

\newcommand{\ket}[1]{{| #1 \rangle}}
\newcommand{\bra}[1]{{\langle #1 |}}

\renewcommand{\vec}[1]{\boldsymbol{#1}}

\newcommand{\fref}[1]{Fig.~\ref{#1}}
\newcommand{\rb}{\vec{r}}

\renewcommand{\eqref}[1]{Eq.~(\ref{#1})}
\newcommand{\fig}[1]{Fig.~(\ref{#1})}
\newcommand{\p}{\boldsymbol{\wp}}
\newcommand{\hp}{\hat{\boldsymbol{\wp}}}
\def \hH{\hat{H}}
\def\L{\mathcal{L}}
\def\hge{\hat{\sigma}^{ge}}  
\def\heg{\hat{\sigma}^{eg}}  
\def\hee{\hat{\sigma}^{ee}}  

\def\hte{\hat{\sigma}^{te}}
\def\het{\hat{\sigma}^{et}} 

\def\sabs{\sigma_{\rm abs}}
\def\ssc{\sigma_{\rm sc}}
\def\rhoee{\rho_{ee}}
\def\rhogg{\rho_{gg}}
\def\rhoeg{\rho_{eg}}
\def\rhoge{\rho_{ge}}
\def\rhoBB{\rho_{\rm BB}}
\def\rhogB{\rho_{g \rm B}}
\def\rhoBg{\rho_{\rm{B}g}}
\def\dotrhogg{\dot{\rho}_{gg}}
\def\dotrhogB{\dot{\rho}_{g \rm B}}
\def\dotrhoge{\dot{\rho}_{ge}}
\def\dotrhoee{\dot{\rho}_{ee}}
\def\dotrhoBB{\dot{\rho}_{\rm BB}}
\def\dotrhoDD{\dot{\rho}_{DD}}

\def\tJ{\tilde{J}}
\def\tG{\tilde{\Gamma}}
\def\GD{\Gamma_{\rm D}}
\def\GT{\Gamma_{\rm T}}
\def\Go{\Gamma_{0}}

\def\Heff{\hat{H}_{\rm eff}}
\def\Jcal{\mathcal{J}}
\def\GthB{\Gamma_{\text{B}}^{\text{th}}}

\def\hge{\hat{\sigma}^{ge}}  
\def\heg{\hat{\sigma}^{eg}}  
\def\hee{\hat{\sigma}^{ee}}

\def\hte{\hat{\sigma}^{te}}
\def\het{\hat{\sigma}^{et}}

\def\hE{\hat{E}}

\def\J{\tilde{J}}

\def\ssc{\sigma_{\rm sc}}
\def\sabs{\sigma_{\rm abs}}

\def\tJ{\tilde{J}}
\def\tG{\tilde{\Gamma}}

\graphicspath{{Figures_Dephasing_Ring/}{Figures_manuscript/}}

\begin{document}
\title{%Decoherence-Enhanced Single-Photon Absorption in a \\ Nanoring of Dipole-Coupled Quantum Emitters
Environment-Enhanced Single-Photon Absorption \\ in a Nano-Ring of Dipole-Coupled Quantum Emitters}%Decoherence Enhancement of Single Photon Absorption \\ in a Nanoring of Dipole-Dipole Coupled Quantum Emitters}
\author{Eric S\'anchez-Llorente}
\affiliation{Departament de F\'{i}sica Qu\`{a}ntica i Astrof\'{i}sica and Institut de Ci\`{e}ncies del Cosmos,\
Universitat\ de\ Barcelona,\ Mart\'{i}\ i\ Franqu\`{e}s\ 1, E-08028\ Barcelona,\ Spain.}
\affiliation{Institut de Ciències del Cosmos, Universitat de Barcelona,
Martí i Franquès 1, E-08028 Barcelona, Spain}

\author{Helmut Ritsch}
\affiliation{Institut für Theoretische Physik, Universität Innsbruck, Technikerstraße 21a, A-6020 Innsbruck, Austria}

\author{Maria Moreno-Cardoner}
%\author{Maria Moreno-Cardoner}
\affiliation{Departament de F\'{i}sica Qu\`{a}ntica i Astrof\'{i}sica and Institut de Ci\`{e}ncies del Cosmos,\
Universitat\ de\ Barcelona,\ Mart\'{i}\ i\ Franqu\`{e}s\ 1, E-08028\ Barcelona,\ Spain.}
\affiliation{Institut de Ciències del Cosmos, Universitat de Barcelona,
Martí i Franquès 1, E-08028 Barcelona, Spain}
%\affiliation{Departament de F\'{i}sica Qu\`{a}ntica i Astrof\'{i}sica and Institut de Ci\`{e}ncies del Cosmos,\
%Universitat\ de\ Barcelona,\ Mart\'{i}\ i\ Franqu\`{e}s\ 1, E-08028\ Barcelona,\ Spain.}
%\affiliation{Institut f\"ur Theoretische Physik, Universit\"at Innsbruck, Technikerstr. 21a, A-6020 Innsbruck, Austria}
\date{\today}

\date{\today}
\begin{abstract}
% last version: input Helmut
Decoherence is mostly detrimental in quantum information and quantum optics applications. However, the interplay between environment-induced incoherent dynamics and unitary evolution can give rise to novel quantum many-body phenomena that can be harnessed as a useful resource. As is well known, in dense subwavelength atomic arrays only a single collective eigenmode in the single-excitation manifold couples strongly to free-space radiation, exhibiting superradiant spontaneous emission. Most of the remaining eigenstates form a manifold of weakly radiative modes, giving rise to long-lived subradiant excitations. Here we demonstrate that populating these subradiant modes via additional decoherence mechanisms, such as dephasing or coupling to phonons, can significantly enhance single-photon absorption in a nanoring of quantum emitters. Such nanoring geometry is particularly appealing due to its unique optical properties and its resemblance to natural light-harvesting complexes, which serve as efficient antennas in photosynthesis. Our findings may shed light on fundamental aspects of energy absorption in nature; despite the much greater complexity of biological systems, they may nonetheless operate according to similar underlying optical principles.

\end{abstract}
\maketitle
\section{Introduction}

Structured subwavelength arrays of quantum emitters constitute a versatile platform for exploring light–matter interactions in the quantum regime \cite{Bettles2016Enhanced,Chang2018Colloquium,Rui2020Subradiant,Sheremet2023Waveguide,Douglas2026ManyBody}. In dense ensembles of quantum emitters, where interparticle distances are of the order of the resonant wavelength or smaller, the quantum emitters couple collectively to the same free-space vacuum modes of the electromagnetic field. This leads to coherent light-mediated dipole-dipole interactions and collective radiation processes \cite{Dicke1954Coherence, Lehmberg1970Radiation, Gross1982Superradiance}. This is in stark contrast to dilute ensembles, where spontaneous emission occurs independently for each emitter. 

Cooperative effects in quantum optics are particularly pronounced in ordered arrays of quantum emitters, where interference between scattered fields can be engineered through geometry. The spatial arrangement of the emitters determines both the strength of coherent dipole–dipole interactions and the collective radiative response \cite{Bettles2015Cooperative, Asenjo-Garcia2017Atomlight, Asenjo-Garcia2017Exponential, Shahmoon2017Cooperative, Ruostekoski2023Cooperative}. In subwavelength arrays, destructive interference will give rise to long-lived collective excitations with strongly suppressed radiative decay rates, commonly referred to as subradiant states \cite{Bettles2016Cooperative, Bettles2016Enhanced, Asenjo-Garcia2017Exponential, Moreno-Cardoner2019Subradianceenhanced}. Generally, collective modes may also exhibit strongly directional emission in a preferred radiation channel while being weakly coupled to other modes of the electromagnetic field \cite{Asenjo-Garcia2017Exponential,Ferioli2021LaserDriven,Jimenez-Jaimes2025Controlling}.

The existence of such subradiant and selectively radiant states has attracted considerable attention due to their potential applications in quantum technologies. Long-lived collective excitations can serve as resources for single-photon storage \cite{Facchinetti2016Storing, Asenjo-Garcia2017Exponential,Cech2023Dispersionless}, the realization of coherent photon–photon interactions and qubit gates implementation  \cite{Paulisch2016Universal,Bekenstein2020Quantum,Moreno-Cardoner2021Quantum, Srakaew2023Subwavelength}, efficient transport \cite{Needham2019Subradianceprotected,Moreno-Cardoner2019Subradianceenhanced,Pal2025Efficient} and the enhancement of quantum metrology protocols \cite{Henriet2019Critical,Zafra-Bono2025Subradiant} % (ho poso a dalt, està relacionat amb photon gates) and the implementation of one- and two-qubit gates for quantum computing using emitter pairs \cite{Paulisch2016Universal}. 
They have also been proposed as building blocks for chiral and topological quantum optics in ordered emitter arrays \cite{Perczel2017Topological, Zhang2019Tunable}. The ability to engineer collective decay through geometry thus provides a versatile tool for quantum information processing and quantum simulation. 

Importantly, geometry-engineered collective radiative states are increasingly accessible in state-of-the-art experiments. Optical tweezer arrays \cite{Endres2016Atombyatom,Barredo2016Atombyatom,Barredo2018Synthetic, Ferioli2021Storage} and optical lattices \cite{Rui2020Subradiant, Srakaew2023Subwavelength} now allow the assembly of ordered emitter configurations with subwavelength spacing and high positional control. Complementary platforms based on semiconductor quantum dots \cite{Tiranov2023Collective} and superconducting qubits coupled to microwave resonators \cite{vanLoo2013PhotonMediated} provide additional routes to engineer collective radiative interactions in solid-state systems. Together, these developments open realistic avenues for implementing and probing the collective phenomena discussed above.

Among the various geometries investigated, rings of quantum emitters exhibit a high degree of symmetry, which gives rise to remarkable collective optical properties \cite{Moreno-Cardoner2019Subradianceenhanced, Holzinger2020Nanoscale}. Owing to their rotational symmetry, ring configurations naturally support collective excitations with well-defined angular momentum. Some of these modes exhibit strongly suppressed coupling to free-space radiation, resulting in highly subradiant states, while others form optically confined modes that resemble cavity-like resonances. Beyond their interest in atomic and solid-state implementations, such ring structures closely resemble natural light-harvesting complexes, including the LH1 and LH2 complexes in purple bacteria \cite{bourne_worster_structure_2019, Montemayor2018Computationala}. This analogy has triggered to engineer biomimetic devices designed for efficient light absorption, excitation transport, and sensing \cite{rebentrostEnvironmentassistedQuantumTransport2009, mattiottiEfficientLightHarvesting2022, Holzinger2020Nanoscale, Moreno-Cardoner2022Efficient, Scheil2023Optical, Pal2025Efficient}.

In practice, real world implementations inevitably involve additional sources of decoherence beyond radiative decay. In particular, coupling to environmental degrees of freedom leads to pure dephasing and thermal fluctuations, which can significantly modify the lifetime and coherence of collective excitations. In atomic systems, such effects may originate from motional fluctuations \cite{eltohfa2025effects}, while in solid-state platforms (including superconducting qubits)  they can arise from fluctuations of the electromagnetic environment. In biomimetic contexts, environmental interactions emulate the protein scaffold surrounding chromophores in organic complexes \cite{bourne_worster_structure_2019, mattioni_design_2021}.

Decoherence is often regarded as a detrimental effect that limits quantum applications. Yet, when combined with collective radiative phenomena, decoherence can acquire a constructive role. As we shall see, controlled dephasing can redistribute population among collective eigenmodes, enhancing the effective coupling to radiatively protected or efficiently absorbing channels \cite{mattiottiEfficientLightHarvesting2022}. This interplay can enable directional excitation transport and, more importantly for the present work, enhance single-photon absorption. These considerations suggest that environmental noise, rather than merely degrading coherence, may be harnessed to optimize light–matter interactions in collective systems.

Motivated by this perspective, we present a systematic study of the interplay between collective radiance and nonradiative decoherence in the process of single-photon absorption by subwavelength rings of quantum emitters. Using the absorption cross section as our central figure of merit, we determine the regimes in which dephasing enhances absorption efficiency and quantify how this enhancement scales with system size and interparticle spacing. Our results demonstrate that collective radiation, when combined with controlled environmental noise, can improve absorption beyond the case of pure coherent dynamics.

The paper is organized as follows. In Sec.~\ref{Sec:Model}, we introduce the model and define the main quantities of interest, including the absorption cross section. In Sec.~\ref{Sec:SingleAtom}, we derive analytical expressions for a single emitter, which serve as a benchmark for the collective response. Sections~\ref{Sec:PureDephasing} and \ref{Sec:ThermalDephasing} analyze the effects of pure and thermal dephasing, respectively, for a nanoring of quantum emitters driven by coherent light in the Dicke limit, corresponding to ring sizes much smaller than the resonant wavelength. In Sec.~\ref{Sec:Size} we go beyond this limit and examine finite-size effects. In Sec.~\ref{Sec:Incoherent} we consider the case of incoherent illumination. Finally, in Sec.~\ref{Sec:Conclusions}, we summarize our main findings.

\section{Model and Definitions}
\label{Sec:Model}
\begin{figure*}
    \centering
    \includegraphics[width=0.75\linewidth]{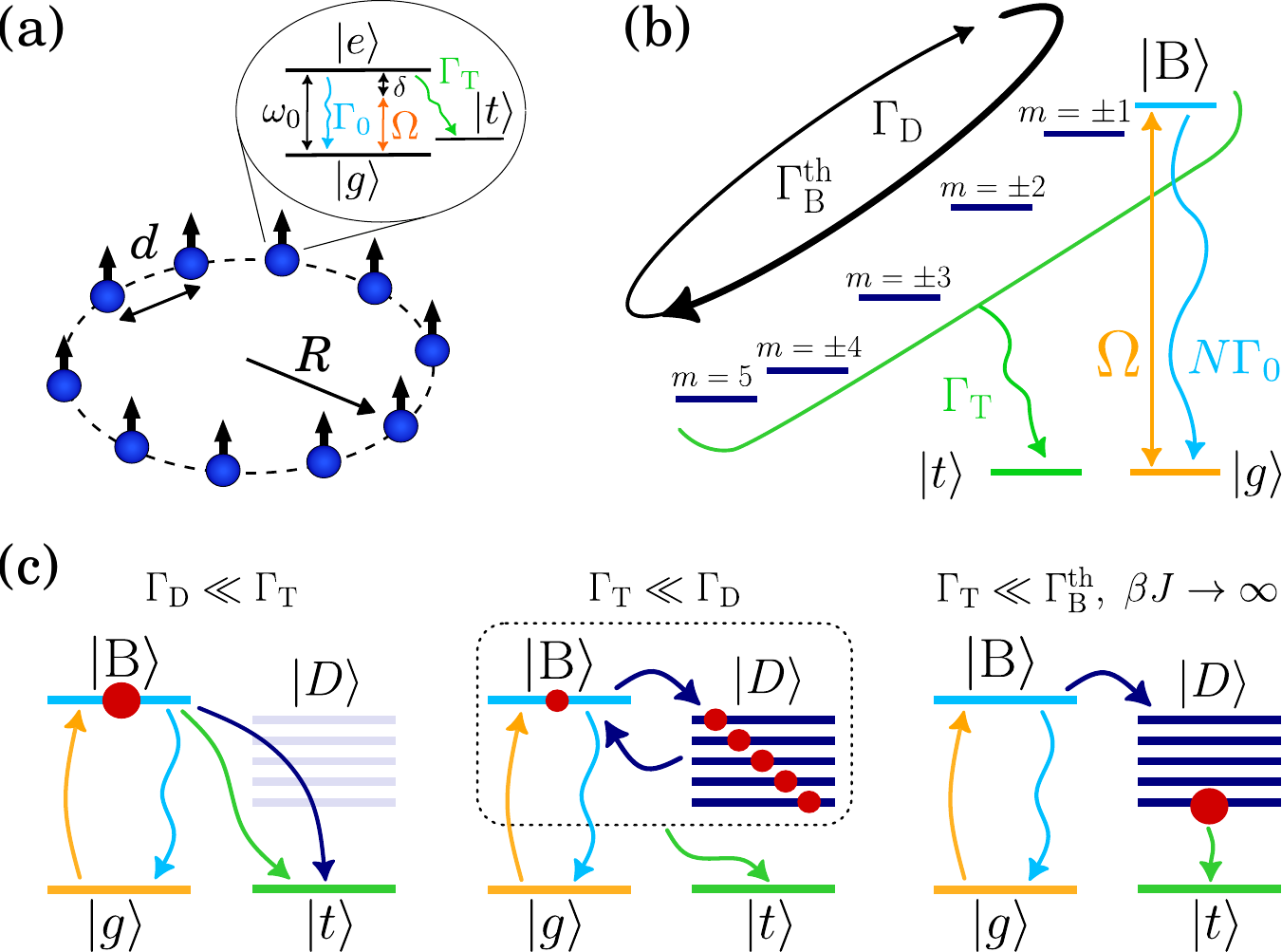}
    \caption{\textbf{(a)} Schematic of the system. A nanoring of three-level quantum emitters ($\ket{g}$, $\ket{e}$ and $\ket{t}$) with radius $R$ and interparticle spacing $d$, coupled through dipole–dipole interactions and driven by a classical field. 
    Black arrows indicate the dipole moments; $\Omega$ and $\delta$ denote the Rabi frequency and detuning of a coherent drive. An emitter can decay from $\ket{e}$ into $\ket{g}$ by spontaneously emitting photons at rate $\Go$, while it irreversibly decays into $\ket{t}$ at rate $\GT$. \textbf{(b)} Level scheme of the single-excitation collective modes of the nano-ring, in the small-volume (Dicke) limit ($R/\lambda_0 \ll 1$). The bright mode ($\rm{B}$) and the ground state ($\rm{g}$) are coherently coupled by the external drive. Local and thermal dephasing, with rates $\GD$ and $\GthB$, respectively, redistribute population among collective modes. \textbf{(c)} Population redistribution in the limiting regimes discussed in the main text. For $\GD \ll \GT$ (or $\GthB \ll \GT$) the excitation remains primarily in the bright mode, and the system behaves as a single emitter with  irreversible decay rate $\GT + \GD$. For $\GT\ll\GD$ the population is uniformly distributed among  collective modes, effectively reducing the radiative decay to $\ket{g}$ by a factor $N$. In the thermal dephasing case, for $\GT\ll\GthB$ and $\beta J \rightarrow \infty$, the population concentrates in the darkest mode, leading to a maximal absorption for any  $\GT\leq N\Go$.} %\textbf{(c)} Sketches of population transfer between the different collective states as a function of the parameters discussed in Sections \ref{Sec:PureDephasing} and \ref{Sec:ThermalDephasing}. The red dots indicate the population distribution among the collective excited states in the steady state. When $\GD\ll\GT$ the population transferred from $\rm{B}$ to the dark modes is negligible and it is rapidly sent to the target state, having an increased average decay to this state. In the opposite limit of $\GT\ll\GD$ the population is evenly distributed among the collective states of the single excitation manifold, all of them with the same decay into the target state. Considering the thermal model of decoherence with zero temperature ($\beta\rightarrow\infty$) the less excited dark mode is the most populated, having a smaller scattering rate and an enhanced absorption.}
    \label{Fig:level-scheme}
\end{figure*}

We consider a regular polygon (from now on nanoring) of $N$ identical two-level quantum emitters trapped at fixed position in the deep subwavelength regime, \emph{i.e.}, the interparticle distances are smaller than the light wavelength $\lambda_0$ associated with the  emitters transition frequency $\omega_0 = c k_0$, being $k_0 = 2\pi/\lambda_0$ the corresponding wave-number. The emitters feature electric dipole moments $\p_i$ ($i = 1,\cdots, N$), with $|\p_i| = \wp$. The distance between two nearest neighbours is denoted by $d$ (see Fig. \ref{Fig:level-scheme}). Throughout this work, we mostly restrict to the case where atoms are transversally polarized to the plane containing the ring. A different dipole orientation can be easily taken into account within the theoretical framework that is introduced later, but it modifies the dispersion relation and the sorting of eigenstates in energy, changing the light absorption efficiency under certain mechanisms, as it will be described later. %For completeness, we present and compare in the Appendix the result for other rotationally symmetric configurations (tangential or radial dipole orientations). {\blue Shall we do this, or remove?}

For interparticle separations comparable to or smaller than the resonant wavelength, the emitters couple to common modes of the electromagnetic vacuum, giving rise to coherent dipole–dipole interactions and collective spontaneous emission. Within the Born–Markov approximation, which is appropriate for many quantum optical platforms, the system dynamics are described by a Lindblad master equation of the form ($\hbar = 1$):
\begin{equation}
\dot{\rho} = -i [ \hat{H} ,\rho ] + \mathcal{L}[\rho].
\label{Eq:MasterEquation}
\end{equation}
In the absence of additional noise or relaxation
channels, the coherent and dissipative parts are entirely determined by the
vacuum-mediated dipole--dipole coupling,
$\hat{H}=\hat{H}_{\rm dd}$ and $\mathcal{L}=\mathcal{L}_{\rm dd}$. 
The coherent contribution is given by 
\begin{equation}
\hat{H}_{\rm dd}
=
\sum_{i\neq j}
J_{ij}\,
\heg_{i}
\hge_{j},
\label{Eq:Hdd}
\end{equation}
while the collective radiative decay is described by \cite{Lehmberg1970Radiation}: 
\begin{equation}
\L_{\rm dd} [\rho] = \frac{1}{2} \sum_{i,j} \Gamma_{ij} \left( 2\hge_j \rho \heg_i - \left\{\heg_i \hge_j, \rho \right\} \right)
\label{Eq:Ldd}
\end{equation}
Here $\hge_j$ ($\heg_j$) is the lowering (raising)  operator between excited and ground state of emitter $j$. The dispersive and dissipative couplings are given, respectively, by $J_{ij} = \hp^{*}_i \cdot \textrm{Re} \bold{G}_{ij} \cdot \hp_j $ and $\Gamma_{ij}=-2\hp^{*}_i \cdot \textrm{Im} \bold{G}_{ij} \cdot \hp_j$, with $\hp_i = \p_i / \wp$ the dipole orientation of the $i$-th atom. Here $\bold{G}_{ij}$ is given by the free space Green's tensor,
\begin{align}
\bold{G}_{ij} &= \frac{3\Gamma_0}{4 }\frac{e^{i k_0 r}}{k_0^3 r^3 } \quad\times\\
&\left[\left(1 -i k_0 r - k_0^2 r^2\right)\mathbb{I}
 +\left(-3+3ik_0 r +k_0^2 r^2\right)\frac{\boldsymbol{r} \otimes \boldsymbol{r}}{r^2} \right]. \notag
\end{align}
In this expression, $\rb = \rb_i -\rb_j$ is the vector connecting dipoles $i$ and $j$, whose modulus is denoted by $r=|\rb|$, and $\Gamma_0 = \wp^2 k_0 ^3 / 3 \pi \epsilon_0$ is the spontaneous emission rate of a single emitter. For the case considered in this work, of dipoles oriented transversally to the ring plane, only the $zz$ component of the tensor is relevant, simplifying into $G_{ij}^{zz} = 3\Gamma_0 e^{i k_0 r}\left(1 -i k_0 r - k_0^2 r^2\right) / 4 k_0^3 r^3$.

The system discussed above is illuminated by coherent light (such as a laser) of very weak intensity and frequency $\omega$. This can be modeled by adding to $\hH$ in Eq.(\ref{Eq:MasterEquation}) the term $\hH_{\rm in} = - \sum_i \left(\Omega_i \heg_i + \rm{h.c.}\right) -\delta \sum_i \hee_i$, with Rabi frequency $\Omega_i \equiv \wp \cdot E^+(\rb_i)/\hbar$ and $E^+(\rb_i) = \avg{\hE^+(\rb_i)}$ the amplitude of the coherent field, that we consider a plane wave of the form $\hE(\rb,t) = \hE^+(\rb) e^{-i \omega t} + \hE^-(\rb) e^{i\omega t}$. Here $\delta=\omega-\omega_0$ denotes the detuning between the external drive frequency $\omega$ and the atomic transition frequency $\omega_0$.\\

{\bf Collective eigenmodes in a nano-ring of quantum emitters.--} 
When the atoms are illuminated by a very weak external field, the dynamics of the system are mostly restricted to the ground state and single excitation manifold, while the population in higher excitation manifolds is negligible. 

With this assumption the term $\hat{\sigma}^{ge}_j\rho\hat{\sigma}^{eg}_i$ in \eqref{Eq:Ldd} can be neglected and the Lindblad master equation corresponding to the dipole-dipole interactions can be written as 
\begin{equation}
    \dot{\rho}=-i\left(\hat{H}_{\rm{eff}}\rho-\rho\hat{H}_{\rm{eff}}^{\dagger}\right),
\end{equation}
where $\hat{H}_{\rm{eff}}=\sum_{ij}(J_{ij} - i \Gamma_{ij}/2) \heg_i \hge_j$ is a non-hermitian effective Hamiltonian. In this single excitation sector, the eigenstates of $\hat{H}_{\rm{eff}}$
%%%%% Aquí acaba la proposta, enllaçaria amb el test que ja hi és %%%%%
%In the single excitation sector, the eigenstates of the effective Hamiltonian $\Heff = (J_{ij} - i \Gamma_{ij}/2) \heg_i \hge_j$ {\blue perhaps i would introduce Heff}%
decay with well defined rate into the ground state, and represent a proper collective basis to describe the dynamics of the system. For the nano-ring geometry considered here, displaying discrete rotational symmetry, collective eigenmodes correspond to spin-waves of the form:
\begin{align}
\ket{m} =  \frac{1}{\sqrt{N}} \sum_{j=1}^N e^{i 2\pi m j /N} \ket{j},
\label{Eq:eigenmodes}
\end{align}
with $\ket{j} = \heg_j \ket{g}$ and well defined angular momentum value $m = 0, \pm 1,\pm 2, \cdots, \lceil \pm (N-1)/2 \rceil$, being $\lceil \cdot \rceil$ the ceiling function. The corresponding collective frequency shifts and decay rates, are then just given by the Fourier transform functions 
\begin{align}
\tJ_m &= \sum_{\ell = j-k} e^{i 2\pi m \ell  / N} J_{jk},\\
\tG_m &= \sum_{\ell = j-k} e^{i  2\pi m \ell  / N} \Gamma_{jk}.
\label{Eq:Collective_shifts}
\end{align}

In this work we focus on a nanoring of quantum emitters transversally polarized with respect to the plane containing the ring. In the first sections, we will restrict to the deep subwavelength regime (\emph{i.e.}, $d/\lambda \ll 1$), also known as the Dicke limit, where the effective Hamiltonian acquires a particularly simple form:
\begin{align}
\Heff = J \sum_{j} (\heg_{j} \hge_{j+1} +  \hge_{j} \heg_{j+1}) - i \frac{\Gamma_0}{2} \sum_{i,j}  \heg_{i} \hge_{j}.
\label{Eq:Heff_Dicke}
\end{align}
The short-range behavior of the dispersive couplings $J_{ij}$ with interparticle distance ($\propto r_{ij}^{-3}$) leads to a coherent term in the Hamiltonian Eq.(\ref{Eq:Heff_Dicke}) that reduces to the tight-binding model, where only first-neighbour couplings are non-zero. This leads to a dispersion relation of the form 
\begin{align}
\tJ_m = J \cos(2\pi m/N),
\label{Eq:DispersionRelation}
\end{align}
with $J =3\Gamma_0/2 k_0^3 d^3 >0$. For positive $J$, the fully symmetric mode $m=0$ corresponds to the maximum of the band and therefore lies highest in energy. In contrast, the dissipative term contains identical couplings between any pair of emitters, and thus, there exists only a single bright mode corresponding to $m=0$, with decay rate $N\Gamma_0$, and $N-1$ states that are dark with $\tilde{\Gamma}_m \approx 0$. As it will be detailed later, this specific order of the states, with a bright mode higher in energy, is crucial to exploit the global thermal bath mechanism for light absorption.\\

%%%
{\bf Light Absorption Efficiency.--}  
To maximize light scattering, we choose the frequency of the external drive $\omega$ in resonance with the single bright mode described before, \emph{i.e.}, $\omega = \tJ_{m = 0}$. Strictly in the Dicke limit, where the remaining modes are perfectly dark, this is the only mode that couples to light. 

In order to model light absorption in the system, we add an irreversible decay channel into an additional trapping state (t), acting as a sink or energy extractor. The role of this state is only to accummulate excitations that are not rescattered into free space, and thus, it does not directly contribute to the dynamics of the remaining degrees of freedom. This additional channel is modeled  as an additional dissipator contributing to $\L$ and defined as:
\begin{align}
\L_{\rm T}[\rho] = \GT \sum_j \left(\hte_j \rho \het_j - \frac{1}{2}\left\{ \hee_j,\rho\right\}\right),
\end{align}
 where $\hte_j$ ($\het_j$) are now the lowering (raising) operator between excited state of emitter $j$ and trapping state. For simplicity, we assume that all sites have the same irreversible decay rate into this state, but a generalization of this model to a site-dependent value of $\GT$ is straightforward. 

Within this model, we quantify the light absorption efficiency through the absorption cross-section. The absorption cross-section per unit beam area defines the probability that an incident photon is absorbed, $\sabs /  A = \dot{n}_{\rm abs}/ \dot{n}_{\rm in}$, where $\dot{n}_{\rm abs}=\GT \rho_{ee}$ and $\dot{n}_{\rm in}$ denote the rates of absorbed and incident photons, respectively. It is convenient to express the absorption cross section in units of the single-emitter scattering cross section $\sigma = 6\pi/k_0^2$. Dividing $\sigma$ by the beam area similarly yields the probability that an incident photon is scattered by a single emitter, $\sigma /  A = \dot{n}_{\rm sc}/ \dot{n}_{\rm in}$, where $\dot{n}_{\rm sc}$ is the single emitter scattering rate. For a plane wave, the incident photon rate per unit area is obtained by dividing the energy flux (energy per unit time and area) by the photon energy, which gives $\dot{n}_{\rm in} = 4|\Omega|^2A/\sigma\Go$. Combining these expressions we obtain:  
\begin{align}
\frac{\sabs}{\sigma}= \frac{ \GT \Gamma_0}{4 |\Omega|^2} \rhoee.
\label{Eq:sigma_abs}
\end{align}
In the following, we focus on the steady-state solution and denote the excited-state population by $\rhoee^{\rm st}$.\\

%%%
{\bf Decoherence mechanisms.--} The main goal of the work is to analyse when the interplay between collective spontaneous emission and dephasing gives rise to an enhanced single photon absorption efficiency. We compare the effect of two different decoherence mechanisms: pure and thermal dephasing. The dynamics described by the two models naturally arises when the electronic excitations are coupled to an external phononic bath (see Appendix \ref{Sec:Appendix_phonons} for a full derivation of the model). Physically, these mechanisms are expected to be present, for instance, in systems of natural light harvesting complexes, arising due to molecular vibrations, or in artificial systems of quantum dots, as a consequence of residual electromagnetic field noise.\\

$\bullet$ {\bf Pure local dephasing model:} We add the dissipator $\L_D$ to the Lindbald $\L$ in Eq.(\ref{Eq:MasterEquation}):
\begin{align}
 \L_{\rm D}[\rho]  =  \GD \sum_j \left( \hee_j \rho \hee_j - \frac{1}{2} \left\{\hee_j, \rho  \right\} \right).
\end{align}
For a $N$-fold rotationally symmetric ring, as the one considered here, it is straightforward to show that the Lindblad $\L_{\rm D}$ acts on the collective eigenmodes basis states Eq.(\ref{Eq:eigenmodes}) as: 
\begin{align}
\dot{\rho}_{mm'} &= \bra{m} \L_{\rm D}[\rho]\ket{m'} =   \GD \left[ \frac{\rho_{ee}}{N} \delta_{m,m'}- \rho_{mm'} \right], \notag\\
\dot{\rho}_{mg} &= \bra{m} \L_{\rm D}[\rho]\ket{g} =   -\frac{\GD}{2} \rho_{mg}.
\label{Eq:LD}
\end{align}
Here $\rho_{mm'} = \bra{m} \rho \ket{m'}$ describes the collective state populations or coherences, depending if $m'=m$ or $m \neq m'$, respectively, and $\rho_{mg} = \bra{m} \rho \ket{g}$. Therefore, \eqref{Eq:LD} describes the effect of local dephasing as an exchange of population between different collective states, and a suppression of all coherences (between different excited states and also with the ground state). Then, it is clear that if only this mechanism is present, the excited state density matrix components evolve towards a maximally mixed state, with all collective modes equally populated, \emph{i.e.}, $\rho_{mm} = \rhoee /N$ ($m = 1,\cdots N$).\\

$\bullet$ {\bf Thermal dephasing model:}
A different mechanism is the coupling with a global thermal bath, with which the system can exchange phonon-like excitations \cite{bourne_worster_structure_2019, mattioni_design_2021, adolphs_how_2006,renger_relation_2002,ishizaki_theoretical_2009,rebentrost_role_2009}. The corresponding dissipator can be written as
%\begin{equation}
   %\mathcal{L}_{th}[\rho]=\sum_{a\neq b}\frac{k_{ab}}{2\pi}\left(\sigma_a^{\dagger}\sigma_b\rho\sigma_b^{\dagger}\sigma_a-\frac{1}{2}\{\sigma_b^{\dagger}\sigma_b,\rho\}\right),
   % \mathcal{L}_{\rm th}[\rho]=\sum_{m\neq m'}k_{m' \rightarrow m}\left(\ket{m}\bra{m'} \rho \ket{m'}\bra{m}-\frac{1}{2}\{\ket{m'}\bra{m'},\rho \}\right),
   %\mathcal{L}_{\rm th}[\rho]=\sum_{a,b}k_{a \rightarrow b}\left(\ket{b}\bra{a} \rho \ket{a}\bra{b}-\frac{1}{2}\{\ket{a}\bra{a},\rho \}\right),
%\end{equation}
\begin{equation}
    \mathcal{L}_{\rm{th}}[\rho]=\sum_{m,m'}k_{m\rightarrow m'}\left(\hat{\sigma}^{eg}_{m'}\hat{\sigma}^{ge}_m\rho \hat{\sigma}^{eg}_m\hat{\sigma}^{ge}_{m'}-\frac{1}{2}\{\hat{\sigma}^{eg}_m\hat{\sigma}^{ge}_m,\rho\}\right),
    \label{Eq:Lth}
\end{equation}
where the indices $m$ and $m'$ run over all possible collective eigenmodes defined in \eqref{Eq:eigenmodes} and $\hat{\sigma}^{eg}_m \equiv \sum_j \braket{j}{m} \heg_j$ ($\hat{\sigma}^{ge}_m$) represent the creation (annihilation) operators of an excitation in mode $m$. These dissipator contributes then to each of the density components as:
\begin{align}
\label{Eq:dynamics_thermal}
\dot{\rho}_{mm'} &= \bra{m} \L_{\rm th} [\rho] \ket{m'} \notag \\
%&= \sum_{b} \left[k_{m\rightarrow b} \rho_{bb} - \frac{1}{2} \left(k_{b\rightarrow m} \rho_{mm'} + k_{b\rightarrow m'} \rho_{m'm} \right) \right]\notag\\
&= \delta_{mm'}\sum_b k_{b\rightarrow m}\rho_{bb}-\frac{1}{2}\sum_b\left(k_{m\rightarrow b}+k_{m'\rightarrow b}\right)\rho_{mm'}\\
\dot{\rho}_{mg} &= \bra{m} \L_{\rm th} [\rho] \ket{g} = -\frac{1}{2}\sum_{b } k_{m\rightarrow b} \rho_{mg}.
\end{align}

%where the $\sigma_c^{(\dagger)}$ annihilate (create) the $c$-th collective mode and 
The transition rate from mode $a$ to mode $b$ is defined as:
\begin{multline}\label{Eq:rates}
    k_{a\rightarrow b} \equiv n(\omega_{ba})\Jcal(\omega_{ba}) + (1+n(\omega_{ab})) \Jcal(\omega_{ab}),
\end{multline}
with $\omega_{ba}=\tJ_b-\tJ_a$ the energy difference between the collective modes $b$ and $a$, $n(\omega) = (e^{\beta \omega}-1)^{-1}$ is the mean number of phonons with frequency $\omega$ following the Bose-Einstein distribution with inverse temperature $\beta$, and $\Jcal(\omega)$ is the spectral density of the bath, which is given by the density of phonon modes weighted by the coupling strength with the system, at frequency $\omega$. 

Following previous work \cite{nalbach_exciton_2011, ishizaki_theoretical_2009, bourne_worster_structure_2019}, %{\red revisar que efectivament fan servir aquesta espectral function}
we will adopt here a spectral function of the Drude-Lorentz form 
\begin{equation}\label{spectral density}
    \Jcal(\omega)=\frac{2 f\omega_c}{\omega^2+\omega^2_c}\omega,
\end{equation}
and restrict to the Ohmic regime for which the function becomes linear with $\omega$, $\Jcal(\omega) \approx \bar{f} \omega$, with $\bar{f} = 2f/\omega_c$. This approximation is valid assuming that all frequencies involved in the problem are smaller than the cut-off frequency $\omega_c$. More refined models, which depend sensitively on the molecular composition and environment \cite{chen_using_2015}, can be employed to describe realistic light-harvesting complexes. However, we expect that the main conclusions of this work are robust with respect to the specific form of the environmental spectral density.

If only thermal dephasing is present in the dissipator, the steady state populations fulfill the detailed balance condition $\rho_{aa} /\rho_{bb} = k_{a\rightarrow b}/k_{b\rightarrow a}$, which for a spectral density such that $\Jcal(\omega) = -\Jcal(-\omega)$, implies the thermal state with populations $\rho_{aa} /\rho_{bb} = e^{-\beta \omega_{ab}}$.
%The transition rates given by \eqref{Eq:rates} depend strongly on the temperature. 

In the high temperature limit ($\beta\rightarrow 0$), where the rate $k_{a\rightarrow b} \approx 2 \bar{f}/\beta = k$ becomes independent of $\omega_{ab}$, the state corresponds to a maximally mixed state, with equal population in each collective mode. We note that in this case, \eqref{Eq:dynamics_thermal} becomes formally equivalent to the population dynamics \eqref{Eq:LD} for the pure dephasing case, when identifying $k = \GD/N$.

In the opposite limit of very low temperature ($\beta\rightarrow\infty$) the rate can be written as $k_{a \rightarrow b}= 2\bar{f}\text{ReLU}(\omega_{ab})$, with $\text{ReLU}(x)$ the rectified linear unit function, being equal to $x$ for positive inputs and $0$ otherwise. Thus, the rate is non-zero and linear with $\omega_{ab}$ for a decreasing energy transition $\ket{a}\rightarrow \ket{b}$, while it is exactly zero for a transition that implies an increase in energy. The thermal Liouvillian then drives the excited state density matrix components into a statistical mixture where only the lowest energy collective state is populated.  %This is exactly what someone would expect intuitively in a situation of zero temperature: the Boltzmann distribution will become huge for the ground state compared with the rest of excited states when the system thermalizes.

\section{Single-Emitter Absorption}
\label{Sec:SingleAtom}

In this section we provide analytical expressions for a single atom which will be used later as a benchmark for the multi-atom case with collective scattering. We consider an atom that couples to light with an electric dipole moment $\wp$ and decays to the ground and trapping states with rates $\Gamma$ and $\GT$, respectively. The atom is driven by a coherent resonant field with Rabi frequency $\Omega$.

The reduced density matrix of the emitter evolves according to the following set of differential equations: 
\begin{subequations}
\begin{align}
\dotrhogg &=  \Gamma\, \rhoee + i \left(\Omega^* \rhoeg -\Omega\rhoge \right),\\
\dotrhoge &= -\frac{1}{2}(\Gamma + \GT + \GD) \rhoge + i \Omega^*(\rhoee-\rhogg), \label{Eq:SingleEmitter_rhoge} \\
\dotrhoee & = -(\Gamma +\GT) \rhoee 
+ i (\Omega \rhoge -\Omega^* \rhoeg).
\label{Eq:SingleEmitter}
\end{align}
\end{subequations}
We find the excited state by setting the time derivatives to be zero ($\dotrhoee  = \dotrhoge = 0$)\footnote{Formally, we should also include the evolution of the target (t) state population, $\dot{\rho}_{tt} = \GT \rhoee$. However, imposing  instead $\dotrhogg=\dot{\rho}_{tt}=0$ would lead to the trivial solution that in steady state all atoms are in (t), as the trapping channel is the only irreversible channel under consideration. We are here instead looking for the asymptotic long-time value of $\rhoee$ before it finally decays into (t).}. We focus in the regime of a very weak driving field ($|\Omega|/\Gamma_0 \ll 1$), where we can approximate $\rhogg^{\rm st} = 1-O(|\Omega|^2) \approx 1$. Plugging this into \eqref{Eq:SingleEmitter_rhoge} we readily find that
\begin{equation}
\rhoee^{\rm 1at,st}(\Gamma) = \frac{4|\Omega|^2}{\Gamma+\GD+\GT}\cdot \frac{1}{\Gamma+\GT},
\label{Eq:rho_ee_1atom}
\end{equation}
which leads to the scattering and absorption cross-sections:
\begin{align}
\frac{\ssc^{\rm 1at}(\Gamma)}{\sigma} &= \frac{\Gamma}{\Gamma+\GD + \GT} \cdot \frac{\Gamma}{\Gamma + \GT},\label{Eq:2L_sc}\\
\frac{\sabs^{\rm 1at}(\Gamma)}{\sigma} &= \frac{\Gamma }{\Gamma+\GD + \GT} \cdot \frac{\GT}{\Gamma + \GT}.\label{Eq:2L_abs}
\end{align}

Interestingly, these expressions can be interpreted as the product of two probabilities. The first term represents the relative excitation probability, i.e., the probability that the atom is promoted from the ground to the excited state in the presence of dephasing and trapping, normalized to the case without these additional decay channels. The second term corresponds to the probability that, once excited, the atom decays through a given channel -- either by spontaneous emission back to the ground state ($\ssc$) or into the trapping state ($\sabs$). In this picture, the two decay pathways, with rates $\Gamma$ and $\GT$, define branching ratios determined by their relative weights with respect to the total decay rate $\Gamma+\GT$. In the absence of trapping and dephasing, the standard result $\ssc = \sigma$ is recovered, as expected.

The absorption cross section of a single emitter with radiative decay rate $\Gamma_0$ is shown in Fig.~\ref{Fig:Absorption_all_models}(a) as a function of the dimensionless parameters $\GT/\Gamma_0$ and $\GD/\Gamma_0$. As follows from~\eqref{Eq:2L_abs}, dephasing always reduces the absorption cross section, since it only acts to suppress the excitation probability. This behavior changes qualitatively in the multi-atom scenario, where emitters couple collectively to the electromagnetic field. As we demonstrate below, in that regime dephasing can instead enhance absorption. This contrast between independent and collective coupling constitutes one of the central results of this work.

On the other hand, $\sabs$ vanishes trivially for $\GT = 0$, since no trapping channel is available. It also tends to zero in the opposite limit $\GT \to \infty$, where strong coupling to the environment suppresses coherent excitation dynamics via the quantum Zeno effect. The absorption cross section therefore reaches its maximum value, $\sabs^{\rm max} = \sigma/4$, at an intermediate point where the two decay rates are equal, $\GT = \Go$. At this optimal point, excitation of the atom is still efficient while the branching ratio toward the trapping channel is maximized. 

\section{Collective absorption under Pure Local Dephasing}\label{Sec:PureDephasing}
We now consider a nano-ring with $N$ atoms in the Dicke (small-volume) limit previously discussed ($d, R \ll \lambda_0$). When illuminating the atoms with a coherent drive, light will only couple to the bright mode. In this case, in absence of dephasing ($\GD=0$) the system behaves as a single atom with decay rates $N \Gamma_0$ and $\GT$, and absorption cross-section identical to $\sabs^{\rm 1at}(N\Go)$ given by \eqref{Eq:2L_abs}. However, dephasing modifies $\sabs$ in  an ensemble of many atoms collectively radiating. 

The reduced density matrix of the emitters evolves according to the following set of differential equations:
\begin{subequations}
\begin{align}
\dotrhogg &=  N \Gamma_0 \, \rho_{\rm BB} + i \sqrt{N} \left(\Omega^* \rhoBg -\Omega\rhogB \right),\label{Eq:rho_gg}\\
\dotrhogB &= [-i(\J_{\rm B}-\delta) -(N\Gamma_0 + \GT + \GD)/2] \rho_{\rm gB} \notag \\ &+ i \sqrt{N}\Omega^*(\rho_{\rm BB}-\rhogg), \label{Eq:rho_gB}\\
\dotrhoBB & = -(N\Gamma_0 +\GD+\GT) \rho_{\rm BB} +\GD \rhoee/N \notag \\
&+ i \sqrt{N} (\Omega \rho_{\rm gB} -\Omega^* \rho_{{\rm B}g} ), \\
%& = -\left[N\Gamma_0 +(N-1)(\GD/N)+\GT\right] \rho_{BB} +(\GD/N)\rho_{DD} + i \sqrt{N} (\Omega \rho_{gB} -\Omega^* \rho_{Bg} )\notag \\
\dotrhoDD  & = -(\GD+\GT) \rho_{DD} + (N-1)\, \GD \rhoee/N. 
%& = -(\GD/N+\GT) \rho_{DD} + (N-1)(\GD/N) \rho_{BB} \notag \\
%\dot{\rho}_{tt} & = \GT \rho_{ee}, 
\label{Eqs1ToyModel}
\end{align}
\end{subequations}
Here $\rm B$ denotes the bright mode, while $D=\left\{1,\cdots,N-1 \right\}$ refers to any of the possible remaining $N-1$ perfectly dark modes, so that $\rhoee = \rhoBB+\sum_{D=1}^{N-1} \rho_{DD}$ is the total excited state population. Similarly as in Sec.\ref{Sec:SingleAtom}, we find $\sabs$ for the steady state when the array is illuminated by a weak driving field ($|\Omega|/\Gamma_0 \ll 1$) and solve the equations imposing $\dot{\rho}_{{\rm Bg}}=\dot{\rho}_{{\rm BB}}=\dot{\rho}_{DD} = 0$ ($\mathrm{D} = \left\{ 1,\cdots,N-1\right\}$) \footnotemark[\value{footnote}].

Again, we approximate $\rhogg = 1 - \mathcal{O}(|\Omega|^2/\Go^2) \approx 1$. in the dynamical equations. Unless otherwise stated, we consider the external drive to be resonant with the bright mode, $\delta = \tilde{J}_{\rm B}$, where absorption is maximized.

Under these conditions, the steady-state equations reduce to the algebraic relations:
\begin{subequations}
    \begin{align}
    (N\Go + \GD+\GT)\rho_{\rm BB} - \frac{\GD}{N} \rhoee
    &= \frac{4 N |\Omega|^2}{\Gamma_{\rm B}^{\rm tot}}, \\
    (\GD+\GT) \rho_{\rm DD} - \frac{\GD}{N} \rhoee&=0 ,
    \end{align}
\end{subequations}
from which the branching ratios are obtained as $ \rho_{\rm BB} / \rhoee= (\GT + \GD/N)/(\GT+\GD)$ and $ \rho_{\rm DD}/\rhoee = (\GD/N) / (\GT + \GD)$. It is  straightforward to find the excited steady state population:
\begin{align}
\rhoee = \frac{4N|\Omega|^2}{N\Gamma_0 + \GD +\GT} \cdot \frac{\GD + \GT}{\GT(N\Gamma_0+\GD+\GT)+ \GD \Gamma_0}.
\label{toy_general1} 
\end{align}

To understand the effect of dephasing in the dynamics, it is instructive to analyze first the limiting cases of $\GD \gg \GT$ and $\GD \ll \GT$.\\

\textbf{Case $\GT \ll \GD\,$.--} In the absence of trapping ($\GT = 0$) and for $\mathcal{L} = \mathcal{L}_{\rm D}$, all collective modes are equally populated in the steady state (see sketch in \fref{Fig:level-scheme}(c)). In this situation, only a single bright mode decays to the ground state at an enhanced rate $N\Go$, while the remaining $N-1$ dark modes are nonradiative. As a result, the total decay is carried exclusively by the bright component, yielding an average decay rate per atom equal to $\Go$, that is, a factor of $N$ smaller than in the case without dephasing. Thus, we find that in steady state the excited-state population is enhanced by a factor of $N$ compared to a single atom with decay $N\Go$, i.e., $\rhoee^{\rm st} = N \rhoee^{\rm 1at, st}$, being $\rhoee^{\rm 1at, st} = 4N|\Omega|^2/(N\Go + \GD)N\Go$. 

In presence of a very small trapping rate ($\GT/\GD \ll 1$) we find an enhanced absorption cross section:
\begin{align}
\left. \frac{\sabs}{\sigma} \right|_{\GT \ll \GD} &=  \frac{N\Gamma_0}{N\Gamma_0 + \GD +\GT} \cdot \frac{ \GT}{\Gamma_0 +\GT}.
\label{Eq:sabs_D}
\end{align}
While the first factor in the previous expression associated with the scattering probability is exactly the same as for the single atom case (decaying with $N\Go$), the second factor, associated with the absorption probability once an atom is excited is enhanced, as there are now $N$ collective modes decaying irreversibly into (t). %Now, there exist $N$ atoms that decay into (t) at rate $\GT$, while the decay rate into (g) per atom has been reduced to $\Gamma_0$. 
In the limit $\GT/\GD \rightarrow 0$, this leads to $\sabs = N\sabs^{\rm 1at}(N\Go)$.\\

\textbf{Case $\GD \ll \GT$.--} 
In the regime $\GD \ll \GT$, the steady-state population of the dark modes remains negligible compared to that of the bright mode (see sketch in \fref{Fig:level-scheme}(c)). Introducing a small but finite $\GD$ effectively opens an additional decay pathway into the trapping state, so that the total trapping rate becomes $\GT + \GD$. In this limit, the absorption cross section $\sabs$ coincides with that of a single effective atom with effective dipole moment $\sqrt{N}\wp$ and enhanced radiative decay rate $N\Gamma_0$, upon replacing $\GT$ by $\GT+\GD$:  
\begin{align}
\left. \frac{\sabs}{\sigma} \right|_{\GD \ll \GT} = \frac{N\Gamma_0}{N\Gamma_0+\GD+\GT} \cdot \frac{\GD+\GT}{N\Gamma_0+\GD+\GT}.
\label{Eq:sabs_T}
\end{align}
Hence, $\sabs$ is maximum with value  $\sabs^{\rm max} = \sigma/4$ now for $N\Go = \GD+\GT$.\\

\begin{figure*}[t]
    \centering
    \includegraphics[width=\linewidth]{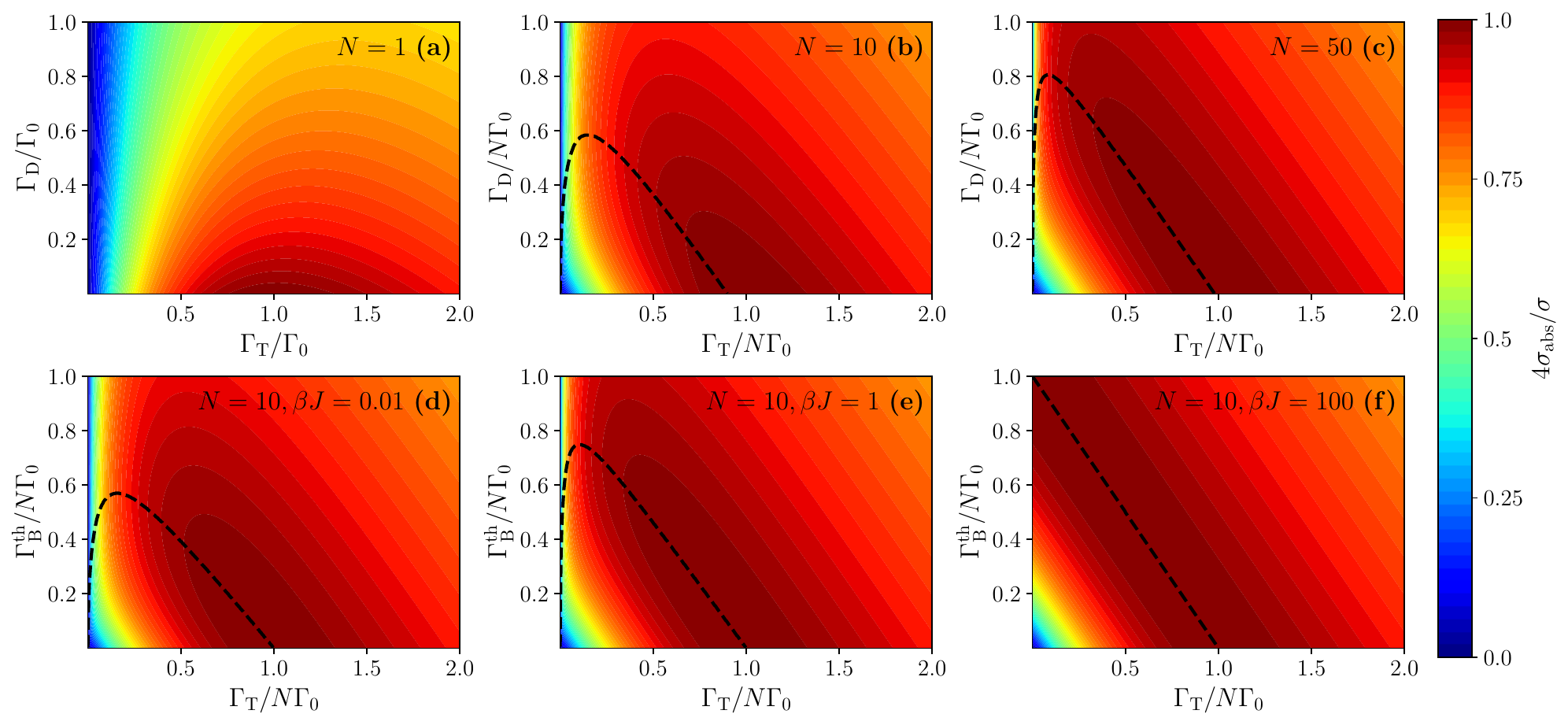}
    \caption{Absorption cross-section $\sabs$ (in units of $\sigma/4$) in the presence of local (top panels) and thermal (bottom panels) dephasing, as a function of $\GT$ and the dephasing rate ($\GD$ or $\GthB$, respectively), normalized to the bright mode spontaneous decay rate $N\Gamma_0$. %Top (bottom) panels are for the local (thermal) dephasing model. 
    {\bf (a)} $N=1$ (single emitter), where $\sabs$ decreases monotonically with $\GD/\Go$ and is maximal at $\GT = \Go$. {\bf (b)} $N=10$ and {\bf (c)} $N=50$. {\bf(d)} $\beta J=10^{-2}$, {\bf (e)} $\beta J=1$ and {\bf (f)} $\beta J=10^2$ for $N=10$. All panels correspond to the Dicke (small-volume) limit. The dashed black line indicates the optimal dephasing rate maximizing $\sabs$ at fixed $\GT/N\Go$. $\sabs$ is always bounded by $\sigma/4$.}
    \label{Fig:Absorption_all_models}
\end{figure*}

\textbf{Arbitrary values of $\GT$ and $\GD$.--} As we have seen, $\GD$ increases the transfer of excitations into dark long-lived modes, but at the same time, it reduces the scattering cross-section and thus the probability that an atom is excited. Therefore, we expect a non-trivial maximum of $\sabs$ as a function of $\GD$. In the general case, the absorption cross-section reads:
\begin{align}
\frac{\sabs}{\sigma}&=  \frac{N\Gamma_0}{N\Gamma_0 + \GD +\GT} \cdot \frac{\GD + \GT}{N\Gamma_0+\GD+\GT+  \GD \Go/\GT},
\label{Eq:sabs_g}
\end{align}
which in the previously discussed regimes $\GT/\GD \ll 1 $ and $\GT/\GD \gg 1$, leads to \eqref{Eq:sabs_D} and \eqref{Eq:sabs_T}, respectively. The absorption cross section \eqref{Eq:sabs_g} of a nanoring with $N=10$ and $N=50$ emitters is plotted in Figs.~\ref{Fig:Absorption_all_models} (b) and (c), respectively, as a function of the dimensionless parameters $\GT' \equiv \GT/N\Gamma_0$ and $\GD'\equiv \GD/N\Gamma_0$. 

We can see that as we increase the value of $N$, $\sabs$ tends to the simple expression 
\begin{equation}
 \lim_{N\rightarrow\infty}\frac{\sabs}{\sigma}=\frac{N\Go }{N\Go+\GT+\GD}\cdot \frac{\GT+\GD}{N\Go+\GT+\GD} ,
 \label{Eq:sabs_local_largeN}
\end{equation}
with an absolute maximum of $\sabs^{\mathrm{max}}=\sigma/4$ when $\GT+\GD = N\Go$. This can be easily seen when writing $\sabs/\sigma$ %=  (\GT'+\GD')/(1+\GT'+\GD')(1+\GT'+\GD'+\GD'/N\GT')$ 
in terms of the parameters $\GT'$ and $\GD'$, and neglecting $\GD'/N\GT'$ for $N\gg 1$. 

% {\blue Proposo això:} We can see that as we increase the value of $N$, $\sabs$ tends to the simple expression 
% \begin{equation}
%  \lim_{N\rightarrow\infty}\frac{\sabs}{\sigma}=\frac{N\Go }{N\Go+\GT+\GD}\cdot \frac{\GT+\GD}{N\Go+\GT+\GD} ,
%  \label{Eq:sabs_local_largeN}
% \end{equation}
% with an absolute maximum of $\sabs^{\mathrm{max}}=\sigma/4$ when $\GT+\GD = N\Go$. {\blue i fins aquí, sense posar l'expressió amb $\GT',\GD'$. Tu prefereixes posar l'expressió amb $\GT',\GD'$ per justificar per què a l'eq (27) no se'n van $\GT,\GD$ quan fem el límit?}{\red Sí, era per justificar això però potser ja es veu...}
% \begin{align}
% \frac{\sabs}{\sigma} &= \frac{\GT'+\GD'}{(1+\GT'+\GD')(1+\GT'+\GD'+\GD'/N\GT')} 
% \end{align}
%which tends to %{\red No estem repetint el mateix %dos cops? No caldria posar-ho en termes de $\GT',\GD'$, no?}
%\begin{equation}
%    \lim_{N\rightarrow\infty}\frac{\sabs}{\sigma}=\frac{\GT'+\GD'}{\left(1+\GT'+\GD'\right)^2}.
%\end{equation}

% $\sabs/\sigma \approx (\GT'+\GD')/(1+\GT'+\GD')^2$,
% for $N\gg 1$. 

In the general case, given a fixed value of $\Go$ and $\GT$, the maximum absorption occurs at:
\begin{equation}
\label{Eq:GD_max}
\GD^{\rm max} =\Gamma_0\sqrt{\frac{N(N-1)\GT}{\GT + \Gamma_0}} - \GT.
\end{equation}
This value is plotted in \fig{Fig:Absorption_all_models} (dashed black line). 
Note that imposing the condition $\Gamma_D^{\rm max} > 0$ in \eqref{Eq:GD_max} yields a constraint on the system parameters for the existence of a nontrivial maximum. This requirement can be written as $\GT(\GT+\Go)/\Go^2 <N(N-1)$, or equivalently,
\begin{align} 
\GT/\Go < N-1.
\label{Eq:Condition_Max}
\end{align}
Therefore, the dimensionless trapping rate $\GT/\Go$ has to be small enough compared to the number of atoms. In fact, for a single atom ($N=1$) and as expected from \eqref{Eq:2L_abs}, dephasing is always detrimental and the maximum absorption occurs  always at $\GD^{\rm max}=0$, regardless of the value of $\GT$. In contrast, when multiple atoms are present ($N\geq 2$), dephasing starts to be beneficial for small enough trapping rates.

At this optimal value of $\GD^{\rm max}$, the absorption cross-section is given by:
\begin{equation}
\label{Eq:sabs_max}
\frac{\sabs^{\rm max}}{\sigma} = \left(\sqrt{\frac{N-1}{N}}+\sqrt{\frac{\Go+\GT}{\GT}}\right)^{-2}.
\end{equation}
It is straightforward to show that this value is always smaller or equal than $\sigma/4$. Specifically, by imposing the above condition given by \eqref{Eq:Condition_Max}, we arrive at $\sabs^{\rm max}/\sigma \leq \left( 1 + N/(N-1) \right)^{-2}$ which is always bounded by $1/4$. This result can be qualitatively understood as it follows. After an excitation is created, and regardless on how it is distributed among the collective eigenmodes, the system either emits a photon and returns back to the ground state with certain probability $p$, or it decays irreversibly into (t) with probability $1-p$. But the probability of absorbing the photon in the first place is always bounded by $p$ (by time reversal symmetry), and therefore, the joint probability of absorption and subsequent conversion into (t) is at most $p(1-p)$, which has a maximum value of $1/4$ when $p=1/2$. 

% When the number of emitters is large we can consider the following limit:
% \begin{equation}\label{Eq:abs_local_limit_N}
%     \lim_{N\rightarrow\infty}\frac{\sigma_{\rm{abs}}}{\sigma}=\frac{N\Gamma_0\left(\GT+\GD\right)}{\left(N\Go+\GT+\GD\right)^2}.
% \end{equation}

%%%%%%% FIGURE 2
% \begin{figure}
% \begin{minipage}[t]{0.45\textwidth}
% \centering
% \includegraphics[width=\textwidth]{sabs_N1.png}
% \end{minipage}
% \begin{minipage}[t]{0.45\textwidth}
% \centering
% \includegraphics[width=\textwidth]{sabs_N10.png}
% \end{minipage}
% \begin{minipage}[t]{0.45\textwidth}
% \centering
% \includegraphics[width=\textwidth]{sabs_N50.png}
% \end{minipage}
% \caption{Absorption cross-section $\sabs$ (in units of $\sigma/4$) versus dimensionless rates $\GT/N\Go$ (trapping) and $\GD/N\Go$ (dephasing), where $\Go$ is the single atom radiative decay rate. {\bf (a)} $N=1$ (single atom). $\sabs$ always decreases with $\GD/\Go$, reaching its maximum at $\GT = \Gamma_0$. {\bf (b)} $N=10$ and {\bf (c)} $N=50$ atoms in the small volume (Dicke) limit. For $\GT/\Gamma_0< N-1$ there exists an optimal dephasing rate $\GD^{\rm max}$ that maximizes $\sabs$ (black dashed line), as collective dark modes are being populated. The absorption cross-section is always bounded by $\sigma/4$ (see main text).}
% \label{Fig:sigma_abs_local}
% \end{figure}

\section{Collective absorption under Thermal Dephasing}\label{Sec:ThermalDephasing}
In the previous section, we demonstrated that dephasing redistributes population into dark modes, thus increasing the effective lifetime of the excitation and enhancing the absorption cross section. However, in that case the redistribution is at most uniform: in the steady state, all collective modes become equally populated. A natural question is whether the absorption can be further enhanced by selectively favoring the occupation of dark modes beyond this uniform limit.

Such selective population can indeed occur in the presence of thermal dephasing when the collective modes have an appropriate energy dispersion, such as the one presented before (see Sec.~\ref{Sec:Model}), with subradiant states lying at a lower energy. In this situation, a thermal environment tends to drive population toward these lower-energy weakly radiative states. As a result, thermal noise does not simply redistribute excitations uniformly, but can preferentially populate dark modes, potentially leading to a further enhancement of the absorption cross section. In the following, we analyze how is this effect depending on the temperature of the bath and the number of emitters.

The master equation \eqref{Eq:MasterEquation} is now defined with $\L=\L_{\rm dd}+\L_{\rm T}+\L_{\rm th}$. The reduced density matrix coefficients are now governed by \eqref{Eq:rho_gg}, together with: 
\begin{subequations}
\begin{align}
\dotrhogB &= [-i(\J_{\rm B}-\delta) -(N\Gamma_0 + \GT + \sum_m k_{\mathrm{B}\rightarrow m})/2] \rho_{\rm gB} \notag \\ &+ i \sqrt{N}\Omega^*(\rho_{\rm BB}-\rhogg),\\
\dotrhoBB & = -(N\Gamma_0 +\sum_{m}%\neq {\rm B}}
k_{{\rm B}\rightarrow m}+\GT) \rhoBB +\sum_{m} k_{m\rightarrow \rm B} \rho_{mm} \notag \\
&+ i \sqrt{N} (\Omega \rhogB -\Omega^* \rhoBg ), \\
\dotrhoDD  & = -\left(\sum_{m}k_{{D} \rightarrow m}+\GT\right) \rho_{DD} + \sum_{m} k_{m \rightarrow D} \rho_{mm}.
\label{Eqs1ToyModel}
\end{align}
\end{subequations}
As before, the index $m$ in the summation runs over all possible states: the bright ($\rm B$) and the dark modes ($D=\{1,\cdots, N-1\}$).

Similarly as in Sec.~\ref{Sec:PureDephasing} we find for a weak driving field which is resonant with the bright mode that the steady state populations fulfill:
\begin{subequations}
\begin{align}
 \Gamma_{\rm B}^{\rm tot} \rho_{\rm BB} - \sum_{m}k_{m \rightarrow \rm B} \rho_{mm} &= \frac{4 { N} |\Omega|^2}{ \Gamma_{\rm B}^{\rm tot}} ,\label{Eq:bright_mode_thermal}\\
\Gamma_{D}^{\rm tot} \rho_{DD} -  \sum_{m }k_{m\rightarrow {D}} \rho_{mm}  &=0, \label{Eq:dark_modes_thermal}
\end{align}
\end{subequations}
where the total decay rate of a given mode $m$ is given by $\Gamma_{m}^{\rm tot} = \tG_m + \GT + \Gamma^{\rm th}_m$, with $\Gamma^{\rm th}_m \equiv \sum_{m'} k_{m\rightarrow m'}$. \\

For an arbitrary value of $\beta$, we can numerically solve this system of linear coupled equations to find the total excited state population $\rhoee = \sum_{m={\rm B},{D}} \rho_{mm}$. %, where $m$ runs over all collective modes (bright and dark modes). 
We then compute the absorption cross section from \eqref{Eq:sigma_abs} and plot the results in \fref{Fig:Absorption_all_models}(d)–(f) as a function of the dimensionless parameters $\GT/N\Gamma_0$ and $\Gamma_{\mathrm{B}}^{\mathrm{th}}/N\Gamma_0$, for different fixed values of $\beta J$.

In this case, $\GthB$ takes the role of $\GD$ in the pure local dephasing case (its explicit dependence on $\beta$ and $\bar{f}$ can be found in Appendix \ref{Sec:Appendix_thermal_B}). Note also that for a fixed value of the inverse temperature $\beta$, $\GthB$ is proportional to the coupling strength $\bar{f}$ between the system and the phonon bath, so in each of the plots the vertical axis scale is proportional to this coupling.

Similarly to the case of pure local dephasing, we find that the absorption cross-section is again bounded by $\sigma/4$. This upper bound is not surprising as the same argument applies here as before: the probability that the  photon is eventually absorbed is always bounded by $p(1-p) \leq 1/4$ ($0\leq p\leq 1$). 

In contrast to the pure local dephasing case, however, the maximum absorption is now attained when $\GT + \GthB \approx N\Gamma_0$. Moreover, the region of parameter space where the absorption remains close to its maximal value expands towards smaller values of $\GT/N
\Go$ as $\beta$ increases (temperature decreases). 
In this sense, increasing $\beta$ plays a role analogous to increasing the number of emitters $N$ in the pure local dephasing case.

\begin{figure}
    \centering
    \includegraphics[width=\linewidth]{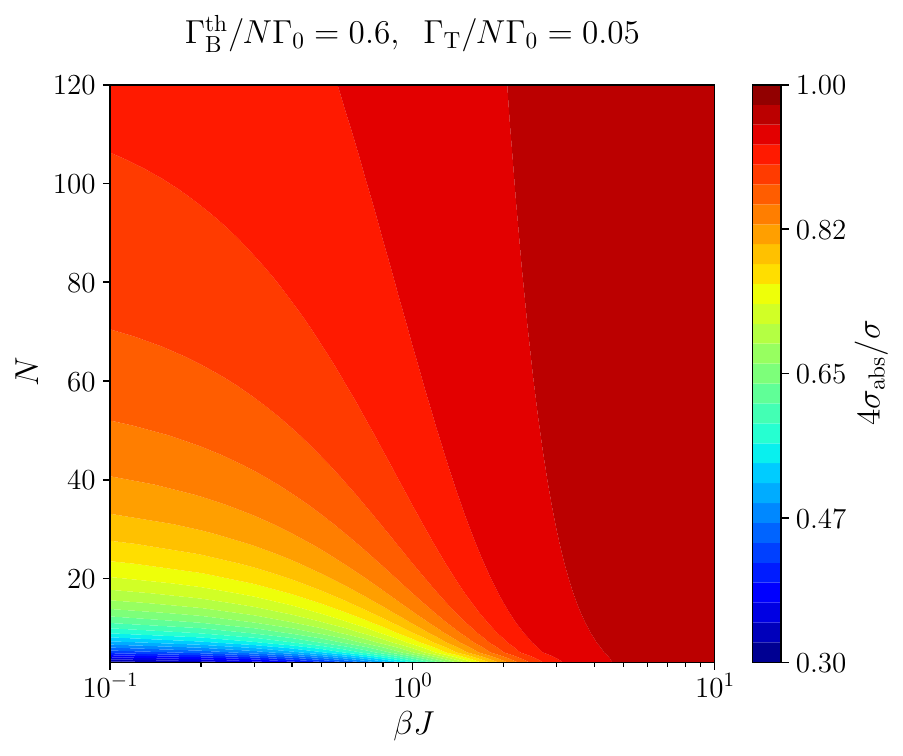}
    \caption{Absorption cross section $\sigma_{\rm abs}$ (in units of $\sigma/4$) in presence of thermal dephasing and in the Dicke (small-volume) limit, at fixed values of $\GT/N\Gamma_0 = 0.6$ and $\GthB/N\Go = 0.05$, plotted versus $N$ and $\beta J$.}
    % The ring radius is fixed to $k_0R=0.001$. {\red També he fet el plot imposant els valors de $\Gamma_m$ i $J_m$ pel cas Dicke, $(N\Go,0,0,...)$ i $(\cos{2\pi m/N})$ respectivament, i obtinc el mateix. Potser ni caldria dir la mida del ring.}}
    %For a specific point in $(\GT/N\Gamma_0,\Gamma^{\text{th}}_{\text{B}}/N\Gamma_0)$-space (in the title) the absorption cross-section $\sigma_{\rm{abs}}$ is measured changing the number of emitters in the ring $N$ and the inverse temperature $\beta J$ while maintaining the radius of the ring fixed to $k_0R=0.001$.}
    \label{Fig:N_equivalent}
\end{figure}

Thus, coupling the system to a low-temperature phonon bath enables the same absorption as in the pure local dephasing scenario, but with a smaller number of emitters. This is illustrated in \fref{Fig:N_equivalent}, where for a particular point in parameter space ($\GT/N\Gamma_0 = 0.05$, $\Gamma_{\text{B}}^{\text{th}}/N\Gamma_0 = 0.6$), we evaluate $\sigma_{\text{abs}}$ as a function of the number of emitters $N$ and the inverse temperature $\beta$. The results show that, along a contour line of constant $\sigma_{\text{abs}}$, the required number of emitters decreases as $\beta$ increases.\\

%%%% FIGURE 4 %%%%
% \begin{figure}[h!]
% \begin{minipage}[t]{0.45\textwidth}
% \centering
% \includegraphics[width=\textwidth]{beta=0.01_with_index_gamma_th.png}
% \end{minipage}
% \begin{minipage}[t]{0.45\textwidth}
% \centering
% \includegraphics[width=\textwidth]{beta=1.0_with_index_gamma_th.png}
% \end{minipage}
% \begin{minipage}[t]{0.45\textwidth}
% \centering
% \includegraphics[width=\textwidth]{beta=100.0_with_index_gamma_th.png}
% \end{minipage}
% \caption{Absorption cross-section $\sabs$ (in units of $\sigma/4$) versus trapping $\GT$ and thermal dephasing $\Gamma_{\rm{B}}^{\rm{th}}$ rates (in units of bright mode spontaneous emission rate $N\Gamma_0$). Each subplot corresponds to a different value of the inverse temperature: \textbf{(a)} $\beta J=10^{-2}$, for which we recover the pure dephasing case result, %where the observed contours match with those obtained using the pure dephasing,
% \textbf{(b)} $\beta J=1$ and \textbf{(c)} $\beta J=10^2$. In all plots $N=10$.}
% \label{Fig:sigma_abs_thermal}
% \end{figure}

%%%% FIGURE 6 %%%%

\begin{figure}[b]
    \includegraphics[width=\linewidth]{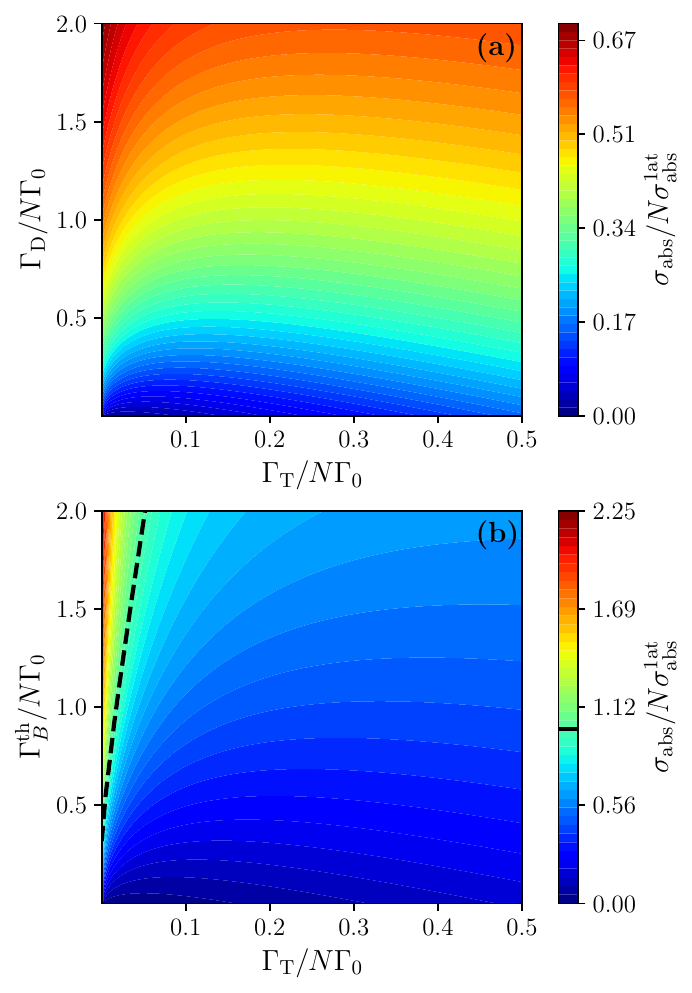}
    \caption{Ratio of the collective absorption cross-section $\sabs$ of a nanoring to that of $N$ independent emitters as a function of trapping ($\GT/N\Go$) and dephasing ($\GD/N\Go$ and $\GthB/N\Go$) rates. {\bf (a)} Pure local dephasing and {\bf (b)} thermal dephasing with $\beta J=1$. 
    In the independent emitter case, each emitter has spontaneous emission rate $\Go$, trapping rate $\GT$, and  
the dephasing rate is taken to be $\GD$ or $\GthB$, respectively. The dashed line marks $1$, corresponding to equal absorption in the collective and independent cases. Results are shown for $N = 10$ in the Dicke (small-volume) limit.}
    \label{Fig:Ratios-bis}
\end{figure}

%%% Ratios
\textbf{Collective absorption enhancement compared to independent atoms.--} %{\blue He posat una mica més clar com agafem $\GD$ pel cas thermal.} The collective enhancement of light absorption in the presence of the phonon bath can be quantified by comparing the absorption cross section $\sabs$ to that of $N$ independent emitters. %Specifically, we consider the ratio between the collective absorption and the total absorption obtained when the same number of emitters decay independently with spontaneous photon emission rate $\Go$ and local dephasing given by $\GD$ or 
The collective enhancement of light absorption in the presence of thermal dephasing can be quantified by comparing $\sabs$ to the total absorption cross-section of $N$ independent emitters. As a reference, we use the single-emitter absorption cross-section with spontaneous emission rate $\Go$ and irreversible decay $\GT$, denoted by $\sabs^{\rm 1at}(\Go)$ and given in \eqref{Eq:2L_abs}. In Fig.~\ref{Fig:Ratios-bis}, we plot the ratio $\sabs / N \sabs^{1\rm{at}}(\Go)$ for both decoherence models. In the case of pure local dephasing, the independent emitters are assumed to have the same dephasing rate $\GD$ as in the collective system. For thermal dephasing, we set $\GD = \GthB$ in \eqref{Eq:2L_abs}, ensuring that independent emitters experience the same effective environmental coupling as in the collective case.
%The absorption cross section of a single emitter with spontaneous photon emission rate $\Gamma_0$ and irreversible decay $\GD$, $\sigma_{\text{abs}}^{1\rm{at}}(\Gamma_0)$, is given by \eqref{Eq:2L_abs}. In Fig \ref{Fig:Ratios-bis} we plot the ratio $\sabs / N \sigma_{\text{abs}}^{1\rm{at}}(\Gamma_0)$ for both models of decoherence. In the case of pure local dephasing the same value of $\GD$ is considered as in the collective case, whereas for the thermal dephasing case we take $\GD = \Gamma_{\rm{B}}^{\rm{th}}$ in \eqref{Eq:2L_abs}). 

For the pure local dephasing case [\fig{Fig:Ratios-bis} (a)], at fixed value of $\GT/N\Go$, the ratio always increases with increasing $\GD/N\Go$. In the limit of $\GD /N\Go \rightarrow\infty$ the ratio reaches one, %meaning that the nano-ring can absorb light into the target state as good as the set of $N$ independent emitters or worse. "sounds less negative: " 
indicating that the nanoring performs equivalently to $N$ independent emitters in transferring population to the target state. Physically, strong dephasing suppresses coherent correlations between emitters and collective interference, such that the absorption efficiency approaches the independent-emitter limit. 

\begin{figure}
    \centering
    \includegraphics[width=\linewidth]{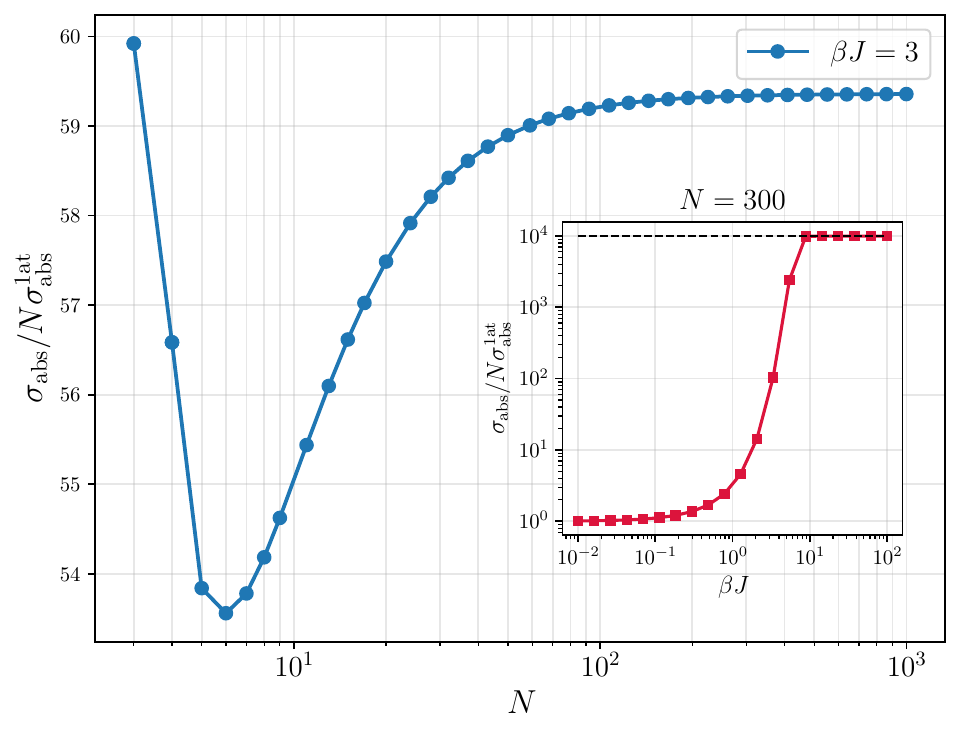}
    \caption{Scaling of the ratio $\sabs/N\sabs^{1\rm{at}}$ with the number of emitters $N$, in the presence of thermal dephasing, for $\beta J=3$. For sufficiently large arrays ($N\gtrsim 100$), the ratio approaches a saturation value that depends on $\beta J$. Inset: Saturation value of $\sabs/N\sabs^{1\rm{at}}(\Go)$, extracted for $N= 300$, as a function of $\beta J$. The gray dashed  line marks the asymptotic low temperature limit from \eqref{Eq:limit_zero_ratio}, scaling as $\sim \Go/\GT$. The trapping rate and coupling strength to the bath are fixed to $\GT/\Go=10^{-4}$ and $\bar{f}=10^{-7}$, respectively.}
   % Scaling of the normalized absorption ratio \(\sabs/(N\sabs^{1\rm at})\) with the number of emitters \(N\), in the presence of thermal dephasing and for \(\beta J=3\). The trapping rate and bath-coupling strength are fixed to \(\GT/\Go=10^{-4}\) and \(\bar{f}=10^{-7}\), respectively. For sufficiently large arrays (\(N\gtrsim 100\)), the ratio approaches a saturation value that depends on \(\beta J\). Inset: Saturation value of \(\sabs/(N\sabs^{1\rm at})\), extracted for \(N\gtrsim 300\), as a function of \(\beta J\). The gray dashed line marks the asymptotic limit of Eq.~\eqref{Eq:limit_zero_ratio}, scaling as \(\sim \Go/\GT\). Other parameters are as in the main panel.  %\blue{Nova proposta de Figura 5, és la que et vaig passar foto ahir. He anat probant diferents valors de $\beta J$ i la que dona un comportament més ``visual'' és $\beta J=3$. Per valors més petits és queden en valors molt baixos sense oscil·lar molt i per valors més alts surten algunes coses rares. $\bar{f}=10^{-7}$, $k_0d=0.001$, $\GT/\Go=10^{-4}$.}} 
    %The interparticle distance is fixed to $k_0d=0.001$. 
    \label{Fig:line_plot}
\end{figure}

In stark contrast, in presence of thermal dephasing [\fig{Fig:Ratios-bis} (b)], the ratio $\sabs / N \sabs^{1\rm{at}}(\Go)$ can surpass one in the regime $\GT \ll \Gamma_{\rm{B}}^{\rm{th}}$. In \fref{Fig:line_plot} we plot the same ratio versus emitter number $N$ for a particular value of  $\beta J = 3$. We observe that as $N$ increases this ratio saturates to a constant value, which increases with inverse temperature $\beta J$, as shown in the inset of the same figure.  Moreover, as we show in the following, for $\beta J \rightarrow \infty$, this ratio scales as $\sabs / N \sabs^{1\rm{at}}(\Go) \sim \Go/\GT$, and thus, it arbitrarily increases when decreasing the trapping rate.\\

%Interestingly, we can derive the expression  in the limit $\beta J \rightarrow \infty$ (zero temperature limit) for which the ratio reduces to, and it arbitrarily increases as $\GT/\Go$ decreases. Moreover, the ratio saturates to a constant value as $N$ increases.
%, as shown in the same figure. 

\textbf{Zero temperature limit.--} In the zero temperature limit ($\beta J\rightarrow\infty$) a transition between states with increasing energy is always forbidden, as the rates $k_{a\rightarrow b}$ vanish for $\omega_{ab}\leq0$. Therefore, \eqref{Eq:bright_mode_thermal} for the bright state population, which is higher in energy, decouples from the other set of equations (since $k_{D \rightarrow {\rm B} } = 0$, for any dark state $D = \left\{ 1, \cdots, N-1 \right\}$), leading to: 
\begin{equation}\label{Eq:bright_mode_population_thermal}
    \rho_{\rm{BB}}=\frac{4 N\left|\Omega\right|^2}{(\Gamma_{\text{B}}^{\text{tot}})^2}=\frac{4N\left|\Omega\right|^2}{(N\Gamma_0+\GT+\Gamma_{\rm{B}}^{\rm{th}})^2}.
\end{equation}
The population of the remaining $N-1$ dark modes can be found iteratively solving the $N-1$ equations \eqref{Eq:dark_modes_thermal} in descending order in energy. Moreover, summing \eqref{Eq:dark_modes_thermal} over all possible dark modes, we obtain the balance equation $\GT \sum_D \rho_{DD} - \GthB \rhoBB = 0$, which directly leads to
\begin{equation}
    \rhoee=\frac{\Gamma_{\rm{B}}^{\rm{th}}+\GT}{\GT}\rho_{\rm{BB}}.
    \label{Eq:InfiniteT_rhoee}
\end{equation}

%The total population in the single excitation subspace can be obtained by adding the contributions from the dark modes,
%\begin{align}
%    \sum_m\Gamma_m^{\rm{tot}}\rho_{mm}&=\sum_{m}\sum_{m'}k_{m'\rightarrow m}\rho_{m'm'}+\sum_{m}k_{\text{B}\rightarrow m}\rho_{\rm{BB}}\nonumber\\
%    &=\sum_{m'}\Gamma_m^{\rm{th}}\rho_{m'm'}+\Gamma_{\rm{B}}^{\rm{th}}\rho_{\rm{BB}}.
%\end{align}
%As $\Gamma_m^{\rm{tot}}=\GT+\Gamma_{m}^{\rm{th}}$ for the dark modes, after cancellations on both sides, one can find
%\begin{equation}
%    \rhoee=\frac{\Gamma_{\rm{B}}^{\rm{th}}+\GT}{\GT}\rho_{\rm{BB}}.
%\end{equation}
% This result is general for any value of the temperature, but is for the limit $\beta\rightarrow\infty$ where the bright mode population is decoupled from those from the other modes.

Using \eqref{Eq:InfiniteT_rhoee} together with  \eqref{Eq:bright_mode_population_thermal} in \eqref{Eq:sigma_abs} we find the simple expression for the absorption cross-section:
\begin{equation}\label{Eq:abs_limit_zero_temperature}
    \left.\frac{\sigma_{\rm{abs}}}{\sigma}\right|_{\beta J\rightarrow\infty}=\frac{N\Go}{N\Go+\GT+\GthB}\cdot \frac{\GT + \GthB}{N\Gamma_0+\GT+\Gamma_{\rm{B}}^{\rm{th}}},
\end{equation}
in agreement with \fig{Fig:Absorption_all_models}(f). Interestingly, this expression closely resembles the result obtained for the local dephasing model \eqref{Eq:sabs_local_largeN} in the limit $N\to\infty$, after replacing $\GD$ by $\GthB$.

%It also corresponds with \eqref{Eq:sabs_local_largeN} for the local dephasing model setting $\GD=\Gamma_{\rm{B}}^{\rm{th}}$ and the limit of large $N$, indicating the connection between the two models and the role of temperature as a resource saving to obtain the same absorption efficiency.
% It also corresponds with the limit of infinite emitters in the nano-ring from the local dephasing model, \eqref{Eq:abs_local_limit_N}, indicating the connection between the two models and the role of temperature being a resource saving to obtain the same absorption efficiency. 
From \eqref{Eq:abs_limit_zero_temperature} one can further obtain an analytical expression for the ratio between the collective absorption cross-section and the one corresponding to $N$ independent emitters. In the regime $\GT \ll \Go, \GthB$ this ratio tends to:
\begin{align}\label{Eq:limit_zero_ratio}   
\lim_{\substack{N\rightarrow\infty }}\left.\frac{\sigma_{\rm{abs}}}{N\sigma_{\rm{abs}}^{1\rm{at}}(\Go)}\right|_{\beta J\rightarrow\infty} &=\frac{\Go(\GthB)^2}{\GT \left(N\Go+\GthB\right)^2} \notag \\ &= \frac{\Go}{\GT} \cdot\left(\frac{2 \bar{f} J}{\Go + 2 \bar{f} J}\right)^2,
\end{align}
where in the last equation we have used that $\GthB = 2\bar{f} J N$ (see  \eqref{thermal_decay_expression} in Appendix \ref{Sec:Appendix_thermal_zero}), i.e., that the thermal dephasing rate scales extensively with the number of emitters. As a consequence, in the presence of thermal dephasing the collective absorption cross section can exceed that of $N$ independent emitters by an arbitrarily large factor as the ratio $\GT/\Go$ is reduced, as shown in \fig{Fig:line_plot}.

\section{Finite-Size Effects in Ring Geometries}\label{Sec:Size}

Up to now we have considered the idealized case where all emitters are contained in a very small volume ($d\ll \lambda_0$), therefore coupling homogeneously to the electromagnetic field. In this regime, commonly referred to as the Dicke limit, there exists only a single bright mode with enhanced decay rate $N\Gamma_0$, and $N-1$ perfectly dark modes with vanishing decay rate. 

We now move beyond this limit and consider a ring of finite spatial extent. In free space, the coupling to the electromagnetic field is no longer homogeneous, and the dark modes acquire a small finite decay rate. 

In the limit of large interparticle separation $d/\lambda_0$, the emitters behave independently and the absorption cross section approaches that of $N$ uncorrelated emitters, namely, $\sabs = N\sabs^{\rm 1 at}(\Go)$. However, for intermediate values of $d/\lambda_0$, the behavior of $\sabs$ is nontrivial. On the one hand, the finite decay rate of subradiant modes reduces the excited-state population and thus $\sabs$. On the other hand, these modes acquire a finite overlap with the driving field, which can increase their population and enhance $\rho_{ee}$ and $\sabs$. The competition between these two effects determines the absorption efficiency in the finite-size regime.

To assess the persistence of the enhancement mechanism discussed above, we analyze the dependence of $\sabs$ on the ratio $d/\lambda_0$. The system dynamics are governed by equations analogous to \eqref{Eq:bright_mode_thermal}, but now each collective mode $m$ is driven with amplitude $\Omega_m = \sum_i \braket{m}{i}\Omega(\rb_i)$. In the steady state the mode populations fulfill (see Appendix \ref{app:thermal_weak_drive} for details):
\begin{align} 
 &\Gamma_{m}^{\rm tot} \rho_{mm} - \sum_{ m'}k_{ m' \rightarrow m} \rho_{ m'm'} = \frac{4 |\Omega_m|^2\Gamma_m^{\text{tot}}}{ (\Gamma_{m}^{\rm tot})^2+4(\tilde{J}_m-\delta)^2}.
 \label{Eqs:finite_size}
\end{align}
For concreteness, we consider a plane-wave drive propagating along the $\hat{x}$ direction with transverse linear polarization, $\Omega(\rb_i)=\Omega_0 e^{i k_0 x_i}$. The detuning is chosen to match the frequency shift of the mode that couples most strongly to light, i.e., the brightest mode with maximal decay rate $\Gamma_{\rm B}$. The value of $\Gamma_{\rm B}$ exhibits oscillations as a function of $d/\lambda_0$, reflecting interference of the scattered fields (see \fref{Fig:Gmax_vs_d} in Appendix \ref{Sec:Gmax_vs_d}). 
% This behavior is shown in Fig.~\ref{Fig:level-scheme}(c).

\begin{figure*}[t]
    \centering
    \includegraphics[width=\linewidth]{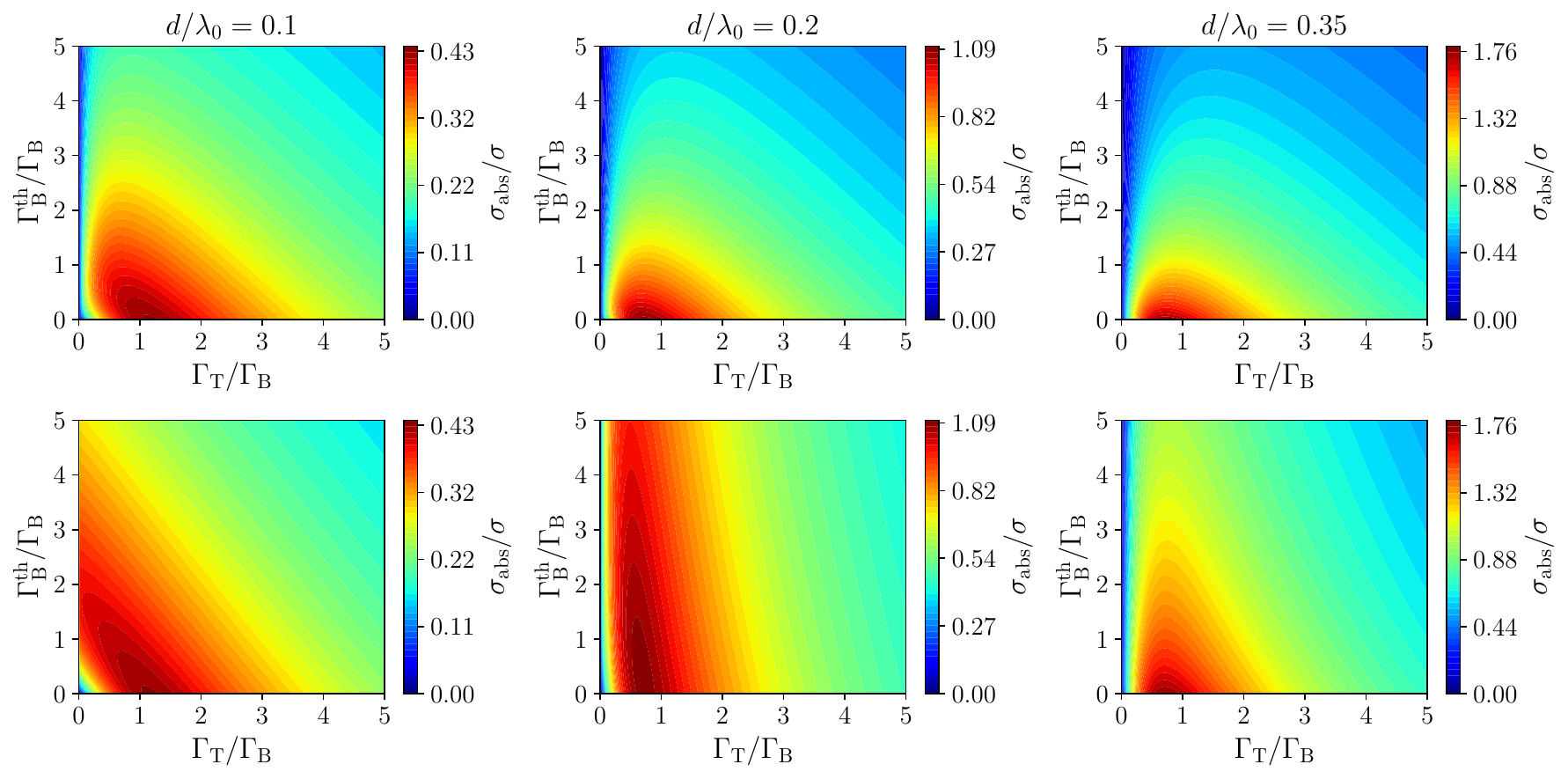}
    \caption{Absorption cross-section $\sabs$ for a finite-size ring. $\sabs$ (in units of $\sigma$) is plotted versus trapping $\GT$ and thermal dephasing $\GthB$ rates, normalized to the bright mode spontaneous decay rate $\Gamma_{\rm B}$, for different values of $d/\lambda_0 = 0.1, 0.2, 0.35$. Top and bottom panels are for $\beta \Go = 0.01$ (high temperature or local dephasing) and $\beta \Go = 10$ (low temperature), respectively. The number of emitters is $N=10$.} 
    % {\blue Aquí estem tenint en compte que $\Gamma_B=\max\{\tilde{\Gamma}_m\}$ i que el detuning és $\delta=\tilde{J}_m(\tilde{\Gamma}_m=\Gamma_B)$.} {\red Es pot apreciar diferència amb el cas $d/\lambda_0=0.2$, no sé si és tema del scaling dels eixos o que :(}}
    \label{Fig:six_plots_detuning}
\end{figure*}

Solving \eqref{Eqs:finite_size} yields the total excited-state population $\rhoee = \sum_m \rho_{mm}$ and, from \eqref{Eq:sigma_abs}, the absorption cross section. In \fref{Fig:six_plots_detuning}, we plot $\sabs$ for $N=10$ emitters as a function of $\GT$ and $\GthB$, normalized by the decay rate of the brightest mode $\Gamma_{\rm B}$. The top panels correspond to $\beta \Gamma_0 = 0.01$ (local pure dephasing regime), while the bottom panels correspond to $\beta \Gamma_0 = 10$ (low-temperature thermal dephasing). Each panel represents a different value of the interparticle spacing, as indicated.

The results show that the enhancement of absorption with increasing $\Gamma_{\mathrm{B}}^{\mathrm{th}}$ at low $\GT$ persists beyond the Dicke limit. As in the fully collective regime, thermal dephasing enlarges the region of parameter space where absorption is maximized, and this effect becomes more pronounced at lower temperatures. However, for $d/\lambda_0 \gtrsim 0.3$, the enhancement due to thermal dephasing vanishes, and the optimal absorption is obtained for $\Gamma_{\mathrm{B}}^{\mathrm{th}} \approx 0$.
 
\begin{figure}[h!]
    \centering
    \includegraphics[width=\linewidth]{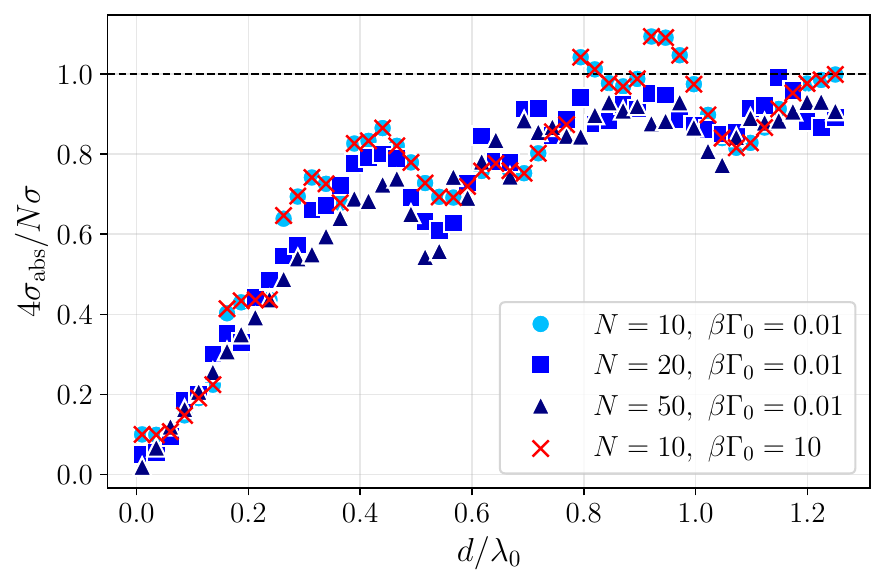}
    \caption{Maximum absorption cross-section $4\sabs / N \sigma$ (optimized over $\GT$ and $\GthB$) for a finite-size ring, as a function of $d/\lambda_0$ at fixed value of $\beta \Go$.  $N=10$, $\beta \Go = 0.01$ (light blue circles); $N=20$, $\beta \Go = 0.01$ (blue squares); $N=50$, $\beta \Go = 0.01$ (dark blue triangles). Results for $N=10$, $\beta\Go=10$ (red crosses) show negligible deviation from the $\beta\Go=0.01$ case with the same number of emitters. The dashed black line indicates the maximum absorption cross-section for $N$ independent emitters.}
    \label{Fig:max_sab_finite_x}
\end{figure}

We also observe, as can be seen in Fig. \ref{Fig:max_sab_finite_x}, that the absolute maximum of $\sabs$ in parameter space increases with $d/\lambda_0$ before eventually approaching the independent-emitter limit. By optimizing $\sabs$ over $\GT$ and $\GthB$ for each spacing, we obtain the maximal achievable absorption as a function of $d/\lambda_0$. This quantity exhibits an oscillatory dependence reflecting interference effects, and asymptotically converges to the expected independent-emitter maximum value, $N\sigma/4$.

\section{Absorption under incoherent illumination}
\label{Sec:Incoherent}
Up to now we have considered coherent driving. We now turn to a different scenario in which the system is illuminated by a field that is spatially coherent but temporally incoherent, such as broadband or thermal radiation (e.g. mimicking sunlight), modeled as a photonic bath following a thermal distribution.  

Accordingly, we set $\Omega = 0$ and include the additional Lindblad contribution
\begin{align}
\mathcal{L}_{\rm inc}[\rho] &= \epsilon N\Gamma_0 \left[ (n+1) S \rho S^\dagger + n S^\dagger \rho S \right.\nonumber\\
 &\left.- \frac{1}{2} \left\{(n+1)S^\dagger S + n S S^\dagger, \rho \right\}\right],
\end{align}
where $S^\dagger = N^{-1/2} \sum_i \heg_i$ creates an excitation in the fully symmetric (bright) mode. The parameter $\epsilon \ll 1$ characterizes the light intensity, while $n$ denotes the mean photon number of the incident field, for instance following a Bose–Einstein distribution at the photon temperature.

The incident photon rate is therefore $\dot{n}_{\rm in} = \epsilon n N\Go $. Analogously to the coherent case, we define the absorption cross section through $\sigma_{\rm abs}/A = \dot{n}_{\rm abs}/\dot{n}_{\rm in}$, yielding
\begin{align}
\frac{\sigma_{\rm abs}^{\rm inc}}{\sigma}
= \frac{\GT}{\epsilon n N \Gamma_0} \,\rhoee.
\label{Eq:sabs_inc}
\end{align}

The reduced density matrix elements now obey
\begin{subequations}
\begin{align}
\dotrhogg &= \epsilon N\Gamma_0 \left[ (n+1) \rho_{\rm BB} - n \rhogg \right] + N\Gamma_0 \rhoBB,\\
\dotrhoBB &= \epsilon N\Gamma_0 \left[ n \rhogg -(n+1) \rhoBB \right] - \Gamma_{\rm B}^{\rm tot} \rhoBB \notag\\
&\qquad{} \qquad{} \qquad{} \qquad{} \qquad{} \qquad{} + \sum_{m}k_{m \rightarrow \rm B} \rho_{mm},
\end{align}
\end{subequations}
while the dark mode populations $\rho_{DD}$ (${D}=\{1,\dots,N-1\}$) evolve as in \eqref{Eq:dark_modes_thermal}.

In the weak-intensity limit ($\epsilon \ll 1$), we approximate $\rhogg = 1 - \mathcal{O}(\epsilon^2)$. Under this condition, the steady-state populations satisfy the same algebraic equations as in the coherently driven case, \eqref{Eq:bright_mode_thermal} and \eqref{Eq:dark_modes_thermal}, upon replacing $4N|\Omega|^2/\Gamma_{\rm B}^{\rm tot} \rightarrow \epsilon N \Go n$. 

The algebraic equations above can be solved numerically for arbitrary values of $\beta J$. As in the coherent case, however, the limits $\beta J \to 0$ (corresponding to pure local dephasing) and $\beta J \to \infty$ (zero-temperature thermal dephasing) admit simple analytical expressions.

In the high-temperature limit $\beta J \to 0$, we obtain:
\begin{align}
\rhoee  &= \frac{\epsilon N\Go n (\GT + \GD)}{\GT (N\Go + \GD + \GT) + \GD \Go}, \\ \frac{\sabs}{\sigma} &= \frac{\GT+\GD}{N\Go + \GD + \GT + \GD \Go/\GT}. 
\end{align}
According to this expression, the absorption cross section increases monotonically with the dephasing rate $\GD$. This behavior can be understood as follows: under broadband illumination, dephasing does not suppress the excitation probability,
% ({\blue Aquest raonament es basa amb el detuning, que té connexió amb la llum incoherent com s'explica després, però no seria millor justificar-ho aquí amb un altre argument? O sino esmentar abans la connexió entre ambdues coses}) 
while it continues to redistribute population from radiative to subradiant collective modes. As a result, $\GD$ enhances the effective trapping efficiency without reducing the probability of photon absorption. This is not the case for a single quantum emitter under incoherent illumination. Consider, in particular, a single emitter that couples to light with the same strength as the bright collective mode, and therefore decays radiatively at rate $N\Go$. For such an emitter the absorption cross section is simply $\sabs^{\rm 1at, inc}(N\Go)= \GT / (N\Go + \GT)$.

We note that in the regime $\GD \gg N\Go, \GT$, where all collective modes are equally populated, $\sabs^{\rm inc}/\sigma = \GT/(\Go + \GT)$, which means that the trapping rate required to reach a given absorption efficiency is reduced by a factor of $N$ compared to the single emitter case with decay $N\Go$. By contrast, in the opposite limit $\GD \ll \GT$, $\sabs^{\rm inc}/\sigma = (\GT+\GD)/(N\Go + \GT + \GD)$ the absorption reduces to that of the single emitter with enhanced irreversible decay rate $\GT + \GD$.

Interestingly, the same expression for the absorption cross section is recovered by averaging the coherent scattering cross section under coherent illumination over all possible detunings (see Appendix~\ref{Sec:Appendix_Incoherent}), highlighting the equivalence between broadband driving and frequency-averaged coherent excitation.

Now we move to the opposite low-temperature limit $\beta J \to \infty$, in which
\begin{align}
\rhoee  &= \frac{\epsilon N\Go n (\GT + \GthB )}{\GT (N\Go+\GT+\GthB)}, \\ 
\frac{\sabs}{\sigma} &= \frac{\GT + \GthB }{N\Go + \GT + \GthB}. 
\end{align}

Again, the absorption cross section increases monotonically with the thermal dephasing rate $\GthB$. In this limit, $\sabs$ coincides with that of a single effective emitter characterized by an enhanced trapping rate $\GT + \GthB$ and a collective radiative decay rate $N\Go$.

\begin{figure}[t]
    \centering
    \includegraphics[width=\linewidth]{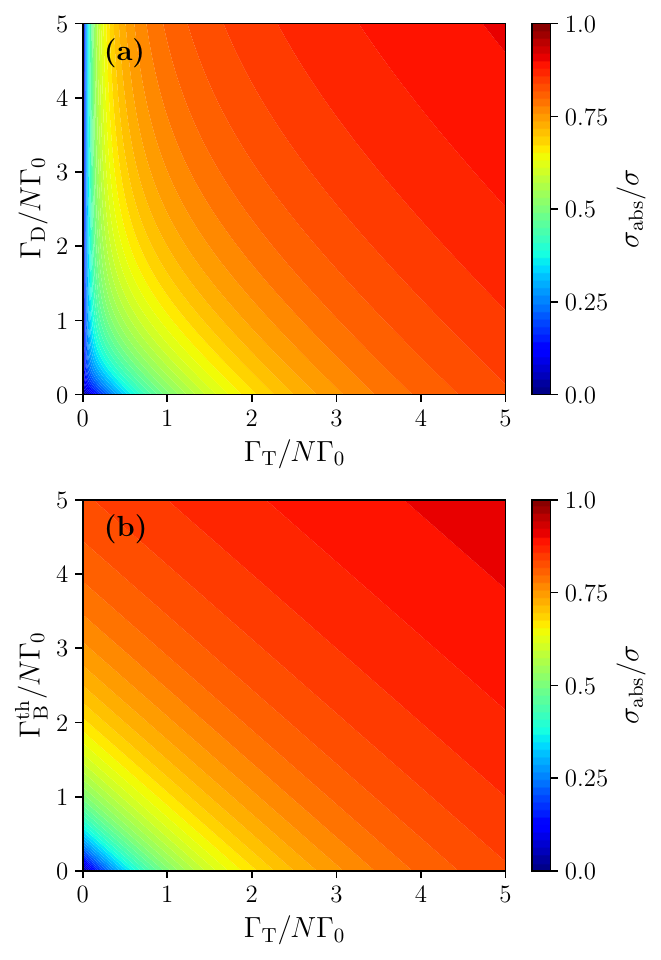}
    \caption{Absorption cross-section $\sigma_{\rm{abs}}$ (in units of $\sigma$) of a nano-ring of quantum emitters under incoherent light, versus traping decay $\GT$ and dephasing $\GD$ or $\GthB$ rates, normalized to the bright mode spontaneous decay rate $N\Go$. {\bf (a)} Local dephasing ($\beta J = 0$). {\bf (b)} Low-temperature thermal dephasing ($\beta J \rightarrow \infty$). All plots are in the Dicke (small volume) limit.}
    \label{Fig:incoherent}
\end{figure}

Figure \ref{Fig:incoherent} shows the absorption cross-section behavior versus $\GT/N\Go$ and $\GD/N\Go$ (or $\GthB/\Go$) in the two limiting regimes $\beta J \rightarrow 0$ and $\beta J\rightarrow \infty$. More generally, we observe that environmental dephasing provides a mechanism to achieve the same absorption efficiency at lower trapping rates $\GT$, regardless of the photon bath temperature.

\section{Conclusions}\label{Sec:Conclusions}
We have shown that collective radiance and environmental dephasing can cooperate to enhance the single-photon absorption cross section in subwavelength nanorings of quantum emitters. Dephasing, typically viewed as detrimental, plays a constructive role here by redistributing population among collective eigenmodes. Depending on the bath temperature, it can populate long-lived subradiant states and thereby increase the effective excitation lifetime. Although the absorption cross-section remains bounded, dephasing enables this bound to be reached for substantially smaller irreversible decay rates into the target state.

For pure local dephasing, increasing the number of emitters enhances absorption by increasing the number of dark modes into which population can be redistributed. For a fixed system size, thermal dephasing produces a similar effect upon lowering the bath temperature, which selectively drives population toward subradiant modes. As a result, thermal dephasing allows the collective absorption to surpass that of $N$ independent emitters, in contrast to the pure local dephasing case.

We further have verified that the main enhancement mechanism survives beyond the Dicke limit: for finite-size rings it persists for interparticle spacings up to $d/\lambda_0 \lesssim0.3$, where dephasing continues to enhance absorption at low $\GT/\Go$. Moreover, under incoherent illumination, dephasing does not reduce the excitation probability and acts only through population redistribution. Consequently, the absorption increases monotonically with both $\GD$ and $\GthB$. 

Together, these results highlight how collective dissipation and environmental noise can be harnessed as resources to optimize light absorption in engineered quantum optical structures. Moreover, we anticipate that similar population-redistribution mechanisms could be used to enhance excitation transport.

\section*{Acknowledgements}
We are deeply grateful to D. E. Chang for proposing the original problem and for generously sharing key early insights and guidance that were crucial in shaping the direction of this work. M.M.-C thanks C. Genes and A. Pal for insightful discussions. E.S.-L thanks Héctor Briongos-Merino and Guillem Lancis Beneyto for helpful discussions. We acknowledge funding from Grant PID2023-147475NB-I00 funded by MICIU/AEI/10.13039/501100011033 and by FEDER, UE. 
E.S.-L. acknowledges support from the 2025 FI STEP 00056 predoctoral grant of the Departament de Recerca i Universitats of the Generalitat de Catalunya, managed by AGAUR (10.13039/501100003030) and co-financed by the European Social Fund Plus (ESF+), as well as from the Quantum Spain project funded by MICIU/AEI/10.13039/501100011033 and NextGenerationEU/PRTR. H. R. acknowledges support from the Austrian Science Fund (FWF) projects 10.55776/FG5
 and the quantA cluster of
excellence10.55776/COE1.

%\bibliography{darkring}
\bibliography{light,radiant_zotero}
\appendix
\clearpage
\section{Microscopic origin of pure local and thermal dephasing models}
\label{Sec:Appendix_phonons}
The dephasing models considered in this work can naturally arise by coupling the system of emitters with a thermal phononic bath, which can be described by an ensemble of harmonic oscillators. In natural light-harvesting complexes, the phononic bath accounts for the vibrational motions of the surrounding molecular and protein environment, whereas in artificial light-harvesting complexes it typically represents the lattice vibrations of the substrate or host material that couples to the electronic excitations. 

The starting point is the Hamiltonian in the single excitation subspace, which can be written as $\hat{H} = \sum_\alpha \omega_{\alpha} \ket{\alpha} \bra{\alpha}+\sum_{q} \omega_q \hat{c}^\dagger_{q} \hat{c}_{q} + \hat{H}_{\rm SB}$. The first term corresponds to the bare system Hamiltonian written in the diagonal collective basis $\left\{ \ket{\alpha} \right\}$, whereas the second term refers to the energy of the vibrational degrees of freedom, described by the creation $c^\dagger_q$ and annihilation $c_q$ operators of a bath excitation. In the so-called Born-Oppenheimer approximation, which applies when there exists a separation of the time scales in the evolution of the system and the bath, the system-phononic bath interaction takes the generic form \cite{mattioni_design_2021}:
\[\hat{H}_{\rm SB} = \sum_{i} \ket{i}\bra{i}  \left[ \sum_{q} \xi_{qi}\left( \hat{c}_q + \hat{c}^\dagger_q\right) \right],\]
where $\xi_{qi}$ is the coupling strength between emitter at position $i$ and the normal bath mode $q$, and it depends on the microscopic details of the bath.

Following standard theory of open quantum systems \cite{breuer_theory_2007,caruso_entanglement_2010}, and if we restrict to the single excitation subspace, the Born-Markov and rotating wave approximations lead to the following Lindblad master equation for the system evolution: 
\[ \dot{\rho} = \sum_{\omega,i,j} \Gamma_{ij}(\omega) \left[ 2 A_j(\omega) \rho A^\dagger_i(\omega)- \left\{ A^\dagger_i(\omega) A_j(\omega), \rho \right\}\right], \]
with $A_i(\omega) = \sum_{\substack{\omega_\beta-\omega_\alpha= \omega}}\ket{\alpha}\braket{\alpha}{i}\braket{i}{\beta} \bra{\beta}$ and $\ket{\alpha}, \ket{\beta}$ denoting the collective eigenmodes of the system. 

The couplings $\Gamma_{ij}(\omega)$ correspond to the Fourier transform of the bath correlation functions and are given by 
\[ \Gamma_{ij}(\omega) = \int_{-\infty}^{\infty} ds e^{i\omega s} \langle B_i^\dagger(s) B_j(0) \rangle, \]
with $B^\dagger_i(t) = \sum_q \xi_{qi} \left( e^{i \omega_q t} \hat{c}_q + e^{-i \omega_q t} \hat{c}^\dagger_q \right)$. Following \cite{mattioni_design_2021} we will then approximate that the spatial correlation part factorizes from this expression, such that we can write $\Gamma_{ij}(\omega) = \eta_{ij} \Gamma(\omega)$, and the value of $\eta_{ij}$ describes the range of spatial correlations of the bath. 

We restrict here to a phononic bath that is in thermal equilibrium, following a Bose-Einstein distribution at temperature $T$, $n(\omega)= (e^{\beta \omega}-1)^{-1}$. After doing some algebra we arrive at the general expression
\[\Gamma_{ij}(\omega) = \eta_{ij} 2\pi \left[ \mathcal{J}(\omega)(1+n(\omega))+ \mathcal{J}(-\omega)n(-\omega)\right],\]
where $\mathcal{J}(\omega) = \sum_{q,i} \xi_{qi}^2 \delta(\omega-\omega_q)$ is the spectral density of the bath. Moreover, we will restrict ourselves to the case of local correlations, $\eta_{ij} = \delta_{ij}$, and define $\Gamma(\omega) = \Gamma_{ii}(\omega)$. 

In the collective basis, we can then rewrite:
\begin{align}
    A_j(\omega) \rho A^\dagger_j(\omega) = \sum \ket{\alpha}\braket{\alpha}{j}\braket{j}{\beta} \bra{\beta}\rho\ket{\beta'} \braket{\beta'}{j} \braket{j}{\alpha'} \bra{\alpha'},
    %_{\substack{\omega_\beta-\omega_\alpha = \omega \\
    %\omega_{\beta'}-\omega_{\alpha'} = \omega 
    %}}
\end{align}
where the summation runs over all states such that $\omega_\beta-\omega_\alpha = \omega$ and $\omega_{\beta'}-\omega_{\alpha'} = \omega$. According to this expression, the dynamics of the diagonal elements of the density matrix $\rho_{\alpha \alpha}$ will only have contributions from $\rho_{\beta \beta'}$ such that $\omega_\beta = \omega_{\beta'}$. Taking into account the form of the eigenstates  \eqref{Eq:eigenmodes}, and that $\braket{\alpha}{j}\braket{j}{\beta}  \braket{\beta'}{j} \braket{j}{\alpha'} \propto \delta_{\alpha-\beta + \beta'-\alpha'}$, together with the fact that the spectrum is anharmonic for the considered case, only the diagonal terms $\rho_{\beta \beta}$ will play a role in the population dynamics. In this situation, the population dynamics are then described by the equation: 
\begin{align}
\dot{\rho} = \sum_{\substack{\omega = \omega_\beta - \omega_\alpha}}\Gamma(\omega) \left[ 2\ket{\alpha} \bra{\beta} \rho \ket{\beta} \bra{\alpha} -
\left\{ \ket{\beta} \bra{\beta}, \rho \right\} \right], 
\end{align}
which is equivalent to \eqref{Eq:Lth}.
%{\red Now I should discuss that taking $\eta_{ij} = \delta_{ij}$ I recover the global model. Perhaps it is simpler from the starting point to consider that the bath does not have spatial correlation? }

%%%%%%%%%%
% \subsection{Evaluation of the absorption cross-sections}
% \label{Sec:Appendix_abs}
% We evaluate the absorption cross section \eqref{Eq:sigma_abs} in presence of a weak driving field $\Omega/\Gamma_0 \ll 1$. To do so, we find the excited state population $\rho^{\rm st}_{ee}$ of the quasi-stationary state by imposing $\dot{\rho}_{ee}=0$ in the master equation \eqref{Eq:MasterEquation}. Note that imposing all derivative components equal to zero (in particular, $\dot{\rho}_{tt} = \dot{\rho}_{gg}=0$) would trivially lead the system into the final state $\ket{t}$, while we are interested in the long time limit transfer of population from $\ket{e_i}$ to $\ket{t}$. 

% Moreover, to the lowest order in $\Omega$, we can always approximate $\rho^{\rm st}_{gg} = 1-O(|\Omega|^2) \approx 1$. Plugging this into \eqref{Eq:MasterEquation} leads to an excited state population which is proportional to $|\Omega|^2$.

% For instance, for a single atom decaying into the ground and trapping states with rates $\Gamma$ and $\GT$, respectively, and in presence of pure dephasing with rate $\GD$, we readily find that 
% \begin{equation}
% \rho^{\rm 1at,st}_{ee}(\Gamma) = \frac{4|\Omega|^2}{\Gamma+\GD+\GT}\cdot \frac{1}{\Gamma+\GT}.
% \label{Eq:rho_ee_1atom}
% \end{equation}
% Replacing this expression in \eqref{Eq:sigma_abs} leads to \eqref{Eq:2L_abs}.

\section{Bright mode thermal dephasing rate}\label{Sec:Appendix_thermal_B}
The thermal decay rate of the bright mode is defined as:
\begin{equation}
    \Gamma^{\rm{th}}_{\rm{B}}=\sum_mk_{\mathrm{B}\rightarrow m},
\end{equation}
where $k_{{\rm B}\rightarrow m}$, given by \eqref{Eq:rates}, is the transition rate from the bright mode to any other mode $m$. Considering the Ohmic limit in the spectral density function and the tight-binding dispersion relation \eqref{Eq:DispersionRelation} we obtain:
\begin{equation}
    \frac{\Gamma^{\rm{th}}_{\rm{B}}}{2\bar{f}J}=\sum_m\frac{1-\cos\left(\frac{2\pi m}{N}\right)}{1-\exp\{-\beta J\left[1-\cos\left(\frac{2\pi m}{N}\right)\right]\}},
\end{equation}
with $m\in\left[-N/2,N/2\right]$. Taking the continuum limit $N\rightarrow\infty$ and defining $x=2\pi m/N$ we can write the previous result as the integral:
\begin{equation}
    \frac{\Gamma^{\rm{th}}_{\rm{B}}}{2\bar{f}J}=\frac{N}{2\pi}\int_{-\pi}^{\pi}dx\frac{1-\cos{x}}{1-e^{-\beta J(1-\cos{x})}}.
\end{equation}
As $|\cos{x}|\leq1$ and $\beta J>0$ the exponential in the denominator is always smaller than unity and the geometric series formula can be applied, leading to:
\begin{equation}
    \frac{\Gamma^{\rm{th}}_{\rm{B}}}{2\bar{f}J}=\frac{N}{2\pi}\int_{-\pi}^{\pi}dx(1-\cos{x})\sum_{n=0}^{\infty}e^{-n \beta J(1-\cos{x})}.
\end{equation}
Bringing the summation out of the integral we find:
\begin{equation}
    \frac{\Gamma^{\rm{th}}_{\rm{B}}}{2\bar{f}J}=\frac{N}{2\pi}\sum_{n=0}^{\infty}e^{-\beta Jn}\int_{-\pi}^{\pi}dx(1-\cos{x})e^{n \beta J\cos{x}},
\end{equation}
where the integral can be written in terms of the modified Bessel functions of First Kind $I_n(z)$. Thus,
\begin{equation}
    \frac{\Gamma^{\rm{th}}_{\rm{B}}}{2N\bar{f}J}=\sum_{n=0}^{\infty}e^{-n\beta J}\left[I_0(n\beta J)-I_1(n\beta J)\right].
\end{equation}
This is a decreasing function with $\beta J$. In the limit of zero temperature $\beta J\rightarrow\infty$ the summation on the right hand side is equal to $1$, which means that $\Gamma^{\rm{th}}_{\rm{B}}=2N\bar{f}J$. In the limit of $\beta J\rightarrow0$ the summation scales as $(\beta J)^{-1}$.

\section{Thermal dephasing rate in the zero temperature limit}\label{Sec:Appendix_thermal_zero}
Moreover, we can find an analytical expression for the thermal dephasing rate of an arbitrary collective mode $m$ in the zero temperature limit. This is now defined as:
\begin{equation}
    \Gamma^{\text{th}}_m=\sum_{m'}k_{m\rightarrow m'},
\end{equation}
with $m'$ running over all other possible collective modes. In the zero-temperature limit $\beta J \rightarrow\infty$, only transitions to lower-energy modes are allowed. Using $k_{a\rightarrow b}=2\bar{f}\,\mathrm{ReLU}(\omega_{ab})$ and the convention $\omega_{mm'}=\tilde{J}_m-\tilde{J}_{m'}$, we obtain
\begin{equation}
    \Gamma^{\text{th}}_m
    =\sum_{m'}2\bar{f}\,\omega_{m m'}
    =2\bar{f}\sum_{m'}\left(\tilde{J}_m-\tilde{J}_{m'}\right),
\end{equation}
where the sum is restricted to modes $m'$ such that $\tilde{J}_{m'}<\tilde{J}_m$.

We now make explicit this restriction using the dispersion relation. In the small-volume (Dicke) limit the collective shifts follow
\begin{equation}
    \tilde{J}_m=J\cos\!\left(\frac{2\pi m}{N}\right),
    \qquad J>0.
    \label{Eq:dispersion_appendix}
\end{equation}
Since $\tilde{J}_{-m}=\tilde{J}_m$, the energies depend only on $|m|$. Moreover, for $|m|\le N/2$ the cosine decreases monotonically with $|m|$ away from $m=0$, and therefore for any $|m|<N/2$,
\begin{equation}
    \tilde{J}_{m'}<\tilde{J}_m
    \quad \Longleftrightarrow \quad
    |m'|>|m|.
    \label{Eq:condition_abs_m}
\end{equation}
For the edge mode $m=\pm N/2$ in the even-$N$ case there are no modes below it in energy.

With this identification, the restricted sum can be split into two contributions:
\begin{equation}
    \Gamma^{\text{th}}_m
    =2\bar{f}\left[
    \sum_{\tilde{J}_{m'}<\tilde{J}_m}\tilde{J}_m
    -\sum_{\tilde{J}_{m'}<\tilde{J}_m}\tilde{J}_{m'}
    \right].
\end{equation}
The first term is independent of $m'$ and simply counts how many states satisfy $\tilde{J}_{m'}<\tilde{J}_m$, i.e.\ how many indices obey $|m'|>|m|$. This yields
\begin{equation}
    \sum_{\tilde{J}_{m'}<\tilde{J}_m}\tilde{J}_m
    =
    \tilde{J}_m \;\max\{0,\,N-2|m|-1\}.
    \label{Eq:first_term_counting}
\end{equation}
The $\max$ function accounts for the darkest states ($m=\pm N/2$ for even $N$ or $m=\pm (N-1)/2$ for odd $N$), for which no lower-energy modes exist.

For the second term, we use again \eqref{Eq:condition_abs_m} and evaluate
\begin{equation}
    \sum_{\tilde{J}_{m'}<\tilde{J}_m}\tilde{J}_{m'}
    =
    \sum_{|m'|>|m|} J\cos\!\left(\frac{2\pi m'}{N}\right).
\end{equation}
This restricted cosine sum admits a closed form, which can be written compactly as
\begin{equation}
    \sum_{\tilde{J}_{m'}<\tilde{J}_m}\tilde{J}_{m'}
    =
    -J\;\max\left\{0,\csc\!\left(\frac{\pi}{N}\right)
    \sin\!\left(\frac{2\pi|m|}{N}+\frac{\pi}{N}\right)\right\}.
    \label{Eq:second_term_closed}
\end{equation}
Again, the $\max$ function ensures the expression also holds for the darkest mode, for which the restricted sum vanishes.

Combining Eqs.~\eqref{Eq:dispersion_appendix}, \eqref{Eq:first_term_counting}, and \eqref{Eq:second_term_closed}, we obtain
\begin{align}
    \frac{\Gamma_m^{\text{th}}}{2\bar{f}J}
    &=
    \cos\!\left(\frac{2\pi m}{N}\right)\;\max\{0,\,N-2|m|-1\}
    \nonumber\\
    &\quad
    +\max\left\{0,\csc\!\left(\frac{\pi}{N}\right)
    \sin\!\left(\frac{2\pi|m|}{N}+\frac{\pi}{N}\right)\right\}.
    \label{thermal_decay_expression}
\end{align}
As stated in the main text, the bright mode satisfies $\Gamma_{B}^{\text{th}}=2\bar{f}JN$, while for the lowest-energy (darkest) modes the thermal decay vanishes, $\Gamma_{m}^{\text{th}}=0$, since there are no states below them in energy.

\section{Spontaneous decay right of the bright mode with a nanoring with finite size}\label{Sec:Gmax_vs_d}
\begin{figure}
    \centering
    \includegraphics[width=\linewidth]{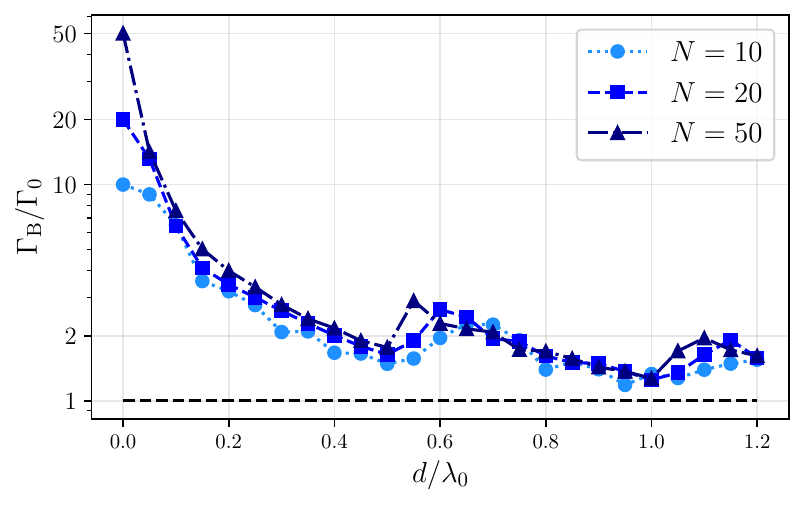}
    \caption{Spontaneous decay rate of the bright mode, $\tG_{\rm{B}}$, as a function of the interparticle distance $d/\lambda_0$ for $N=\{10,20,50\}$ emitters.}
    \label{Fig:Gmax_vs_d}
\end{figure}
In \fref{Fig:Gmax_vs_d} we show how the spontaneous decay rate of the most radiant collective mode varies with the interparticle distance between the emitters in the nanoring. When $d\ll\lambda_0$ we enter in the deep subwavelength regime or Dicke limit where $\tG_{\rm{B}}\approx N\Go$ and the bright mode corresponds to the $|m=0\rangle$ state. As $d$ increases $\tG_{\rm{B}}$ shows an oscillatory behavior due to the changes in the interference pattern of the emitted field. In particular, the decay rate of the permutationally symmetric mode can be reduced, and for certain separations modes with $m\neq 0$ become more radiative than the symmetric one. As the separation between consecutive emitters goes beyond the subwavelength regime, the decay approaches the single atom spontaneous decay rate, $\tG_{\rm{B}}\approx\Go$. This happens when we reach the independent emitters regime, where the effects of the dipole-dipole interactions become negligible.
%, making not only that the permutationally symmetric state decreases its own decay but that other states with $m\neq0$ radiate more than this symmetric mode. 

\section{Derivation of weak-driving equations for the finite-size ring }
\label{app:thermal_weak_drive}

Here we generalize the single-mode weak-driving dynamical equations to the case where several collective modes are excited (finite-size ring), also in presence of a thermal dephasing bath. We work in the single-excitation approximation with basis formed by the ground state \(\ket{g}\) and the collective excited states \(\ket{m}\). The density matrix evolves according to
\begin{align}
\dot{\rho}
&=
-i\left[
(\hH_{\rm eff}+\hH_{\rm in})\rho-\rho (\hH_{\rm eff}^{\dagger} + \hH_{\rm in})
\right] \notag\\
&+
\sum_m \tilde\Gamma_m
\ket{g}\bra{m}\rho\ket{m}\bra{g}
+
\mathcal{L}_{\rm th}[\rho],
\label{eq:appendix_master_equation}
\end{align}
and in the collective basis the effective and input Hamiltonians can be written as:
\begin{align}
\hat{H}_{\rm eff}
&=
\sum_m
\left[
-\delta_m -
i\frac{\tilde\Gamma_m}{2}
\right]
\ket{m}\bra{m}
\notag\\
\hH_{\rm in} &= -
\sum_m
\left(
\Omega_m\ket{m}\bra{g}
+
\Omega_m^*\ket{g}\bra{m}
\right),
\label{eq:appendix_heff}
\end{align}
where $\Omega_m = \sum_i \braket{m}{i}\Omega(\rb_i)$ is the spatially dependent Rabi frequency projected onto collective mode $m$, and $\delta_m \equiv \delta - \tilde{J}_m$ is the detuning with respect to mode $m$. 

From \eqref{eq:appendix_master_equation}, the ground-state population, ground-excited coherences and excited state density matrix elements obey, respectively:
\begin{widetext}
\begin{align}
\dot{\rho}_{gg}
&= \sum_m \tilde\Gamma_m\rho_{mm}
+i\sum_m
\left(
\Omega_m^*\rho_{mg}
-
\Omega_m\rho_{gm}
\right)
\\
\dot{\rho}_{gm}&=
\left(i\delta_m
-
\frac{\Gamma_m^{\rm tot}}{2} \right) \rho_{gm}
+
i\sum_n \left( \Omega_n^*\rho_{nm}
-\Omega_m^*\rho_{gg} \right)
\label{eq:appendix_rhogm}
\\
\dot{\rho}_{mm'}&=
\left[ i
\left(
\delta_m-\delta_{m'}
\right)
-
\frac{\tilde\Gamma_m+\tilde\Gamma_{m'}}{2} -\GT \right]\rho_{mm'}
+i\left( \Omega_m\rho_{gm'}
-
\Omega_{m'}^*\rho_{mg}\right) + 
\delta_{mm'}\sum_b k_{b\rightarrow m}\rho_{bb}
-
\frac{1}{2}
\sum_b
\left(
k_{m\rightarrow b}
+
k_{m'\rightarrow b}
\right)
\rho_{mm'},
\label{eq:appendix_rhommprime}
\end{align}
\end{widetext}
where we have defined the total decay rate (linewidth) of mode \(m\) as
\begin{equation}
\Gamma_m^{\rm tot}
\equiv
\tilde\Gamma_m+ \GT+ \sum_b k_{m\rightarrow b}.
\label{eq:appendix_gamma_tot}
\end{equation}
For the diagonal components, \(m=m'\), this reduces to
\begin{equation}
\dot{\rho}_{mm}
=
-\Gamma_m^{\rm tot}\rho_{mm}
+
\sum_b k_{b\rightarrow m}\rho_{bb}
+
i\left( \Omega_m\rho_{gm}
-
\Omega_m^*\rho_{mg} \right).
\label{eq:appendix_rhomm}
\end{equation}

We now take the weak-driving limit. In this regime the system remains mostly in the ground state,
$\rho_{gg}\simeq 1$, 
while the optical coherences are first order in the drive amplitude, $\rho_{gm},\rho_{mg}=O(\Omega)$, 
and the excited-state populations and coherences are second order,
$\rho_{mm'}=O(|\Omega|^2)$. 
Therefore, to leading order in the drive, the last term in \eqref{eq:appendix_rhogm}, which is proportional to \(\Omega_n^*\rho_{nm}\), can be neglected. At steady state we obtain
\begin{equation}
0
=
i \delta_m \rho_{gm}
-
\frac{\Gamma_m^{\rm tot}}{2}\rho_{gm}
+
i\Omega_m^*.
\end{equation}
Thus,
\begin{equation}
\rho_{gm}^{\rm st.}
=
\frac{i\Omega_m^*}
{i \delta_m -\Gamma_m^{\rm tot}/2},
\label{eq:appendix_rhogm_ss}
\end{equation}
and, equivalently,
\begin{equation}
\rho_{mg}^{\rm st.}
=
\frac{i\Omega_m}
{i \delta_m +\Gamma_m^{\rm tot}/2}.
\label{eq:appendix_rhomg_ss}
\end{equation}

Substituting these weak-driving coherences into \eqref{eq:appendix_rhomm}, the optical source term becomes
\begin{align}
i\left(\Omega_m\rho_{gm}^{\rm st}
-
\Omega_m^*\rho_{mg}^{\rm st} \right)
=
\frac{4|\Omega_m|^2\Gamma_m^{\rm tot}}
{(\Gamma_m^{\rm tot})^2+4(\tilde{J}_m-\delta)^2}.
\label{eq:appendix_source}
\end{align}
Therefore, the steady-state population equation becomes
\begin{equation}
\Gamma_m^{\rm tot}\rho_{mm}
-
\sum_{m'} k_{m'\rightarrow m}\rho_{m'm'}
=
\frac{4|\Omega_m|^2\Gamma_m^{\rm tot}}
{(\Gamma_m^{\rm tot})^2+4(\tilde{J}_m-\delta)^2},
\label{eq:appendix_final_population}
\end{equation}
as in \eqref{Eqs:finite_size}. This is a generalization of the single-mode weak-driving result. The left-hand side describes radiative loss plus thermal depletion of mode \(m\), together with thermal feeding from the other collective modes. The right-hand side is the optical pumping rate into mode \(m\), broadened by the total linewidth \(\Gamma_m^{\rm tot}\).

\section{Broadband illumination (off-resonant light)}\label{Sec:Appendix_Incoherent}

Here we show that incoherent (thermal) illumination yields the same steady-state populations as a coherently driven system after averaging over the detuning $\delta$. Broadband excitation can be viewed as an ensemble of monochromatic drives with uniformly distributed frequencies, such that the incoherent result is recovered by integrating the coherent steady-state response over detuning, provided one identifies the effective incident photon flux via $4|\Omega|^2 / N\Gamma_0 \rightarrow \epsilon n N\Gamma_0$. 

%It is interesting to see that the expressions with thermal light are exactly equivalent to those obtained by driving the system out of resonance and then averaging over all possible frequencies, after replacing the incident rate of photons $4|\Omega|^2 / N\Gamma_0 \rightarrow \epsilon n N\Gamma_0$. 

Dephasing may be interpreted as rapid fluctuations of the transition frequency (equivalently, phase randomization). Under broadband illumination, the excitation probability is no longer sensitive to the detuning of a single monochromatic tone, because the response is effectively averaged over $\delta$. As a result, the detuning-dependent reduction of the extinction is removed, and dephasing primarily affects the dynamics by redistributing population among available states rather than by suppressing coherent excitation.

The situation is particularly relevant in the collective case. For homogeneous local dephasing (i.e., identical dephasing rates on all sites), all collective modes experience the same statistical frequency fluctuations, and the population exchange between collective modes is not impeded by detuning. More generally, for inhomogeneous dephasing one still obtains mode mixing whenever the corresponding overlap factors in the collective basis remain nonzero; however, the effective exchange rates typically decrease as the inhomogeneity increases.

% We can think of dephasing as fast oscillations in the energy levels. Then, it seems natural that, at least for a two-level system, illuminating the system with broadband light will wash any possible effect of this dephasing. Creating an excitation in the first place will become a very  inefficient process, and then, the balance discussed before between the decay rates in the system is modified. Now, since the probability of creating an excitation does not play anymore a role in the absorption cross-section, this montonically increases with $\GT$. 

% The situation is different in the collective case. If we consider a homogeneous local dephasing (dephasing rates are equal for all sites)\footnote{Also in a more general situation where dephasing is not homogeneous, even the different levels oscillate at different frequency, still there is mixing between the exciton components as far as $\sum_{i,j}(\Gamma_i +\Gamma_j)\rho_{ij} \braket{k}{i} \braket{j}{k} \neq 0$. However, I suppose the exchange rate decreases with the inhomogeneities. \rm}., all modes are effectively oscillating with the same frequency, and detuning does not prevent from the exchange between the population modes. 

%More specifically, we can define the average scattering cross-section for a two-level atom:
To make the equivalence explicit, we consider the detuning-integrated scattering cross section of an effective two-level emitter with radiative width $N\Go$:
\begin{align}
\sigma(\delta) = \sigma \frac{(N\Gamma_0)^2}{(N\Gamma_0)^2+4\delta^2}.
\end{align}
Integrating over the detuning yields:
\begin{align}
\avg{\sigma} =  \int \sigma (\delta) d \delta = \frac{\pi}{2} N\Gamma_0 ~\sigma.
\end{align}

Likewise, the extinction cross section under coherent driving,
\begin{equation}
\sigma_{\rm ext}(\delta)\equiv \sigma_{\rm sc}+\sigma_{\rm abs}
= \sigma\,
\frac{N\Gamma_0\left(N\Gamma_0+\GD+\GT\right)}
{\left(N\Gamma_0+\GD+\GT\right)^2+4\delta^2},
\end{equation}
satisfies
\begin{equation}
\int_{-\infty}^{\infty}\sigma_{\rm ext}(\delta)\,d\delta
=\frac{\pi}{2}\,N\Gamma_0\,\sigma,
\end{equation}
showing that detuning integration removes any dependence on $\GD$ and $\GT$ from the overall extinction (excitation) probability.

%with  $\sigma(\delta) = \sigma (N\Gamma_0)^2 / [(N\Gamma_0)^2+4\delta^2]$. The average of the extinction cross-section   
%\[   \sext(\delta) \equiv  \ssc + \sabs = \sigma \frac{N\Gamma_0 (N\Gamma_0 + \GD + \GT) }{(N\Gamma_0+\GD+\GT)^2 + 4\delta^2},\] 
%leads to exactly the same after averaging over all frequencies:
%\[ \avg{\sext} \equiv \int_{-\infty}^{\infty} \sext (\delta) d\delta = \frac{\pi}{2} N\Gamma_0 \sigma,\]
%so that the dependence on $\GT$ and $\GD$ of the probability of creating an exciton has been completely removed. 

The corresponding detuning-integrated absorption cross section is then:
\begin{align}
\avg{\sabs} = \frac{\GD + \GT}{N\Gamma_0 + \GD + \GT+ \GD \Gamma_0/\GT} \avg{\sigma},
\end{align}
which reproduces the same functional dependence as obtained for incoherent illumination (upon the photon-flux identification above).

\end{document}